\documentclass[11pt]{article}

\usepackage[final]{acl}

\usepackage{times}
\usepackage{latexsym}
\usepackage{multirow}
\usepackage{amssymb}
\usepackage{amsmath}
\usepackage{tcolorbox}
\usepackage{orcidlink}

\usepackage[T1]{fontenc}

\usepackage{times}
\usepackage{latexsym}
\usepackage[utf8]{inputenc}
\usepackage{microtype}
\usepackage{booktabs}
\usepackage{xcolor}
\usepackage{tcolorbox}
\tcbuselibrary{skins, breakable}

\usepackage{booktabs}
\usepackage{colortbl}
\usepackage{xcolor}
\usepackage{multirow}

\definecolor{warm1}{HTML}{E8960A}
\definecolor{warm2}{HTML}{F0A820}
\definecolor{warm3}{HTML}{F5BC45}
\definecolor{warm4}{HTML}{F7CC6A}
\definecolor{warm5}{HTML}{F9D98A}
\definecolor{warm6}{HTML}{FAEAA8}
\definecolor{warm7}{HTML}{FBF0C0}
\definecolor{warm8}{HTML}{FCF5D8}
\definecolor{warm9}{HTML}{FEFAEE}
\definecolor{c1frame}{RGB}{52,101,164}   
\definecolor{c1bg}   {RGB}{235,242,252}
\definecolor{c2frame}{RGB}{78,153,78}    
\definecolor{c2bg}   {RGB}{237,250,237}
\definecolor{c3frame}{RGB}{160,82,45}    
\definecolor{c3bg}   {RGB}{252,244,236}
\definecolor{c4frame}{RGB}{100,100,170}  
\definecolor{c4bg}   {RGB}{242,241,252}
\definecolor{c5frame}{RGB}{100,100,100}  
\definecolor{c5bg}   {RGB}{245,245,245}
 \tcbset{
  vocabbox/.style={
    enhanced, breakable,
    fonttitle=\bfseries\small, coltitle=white,
    attach boxed title to top left={yshift=-2mm, xshift=5mm},
    boxed title style={rounded corners, size=small},
    boxrule=0.5pt, arc=3pt,
    left=5pt, right=5pt, top=7pt, bottom=5pt,
    before skip=5pt, after skip=5pt,
  },
  cat1/.style={vocabbox, colback=c1bg, colframe=c1frame,
               boxed title style={colback=c1frame, rounded corners}},
  cat2/.style={vocabbox, colback=c2bg, colframe=c2frame,
               boxed title style={colback=c2frame, rounded corners}},
  cat3/.style={vocabbox, colback=c3bg, colframe=c3frame,
               boxed title style={colback=c3frame, rounded corners}},
  cat4/.style={vocabbox, colback=c4bg, colframe=c4frame,
               boxed title style={colback=c4frame, rounded corners}},
  cat5/.style={vocabbox, colback=c5bg, colframe=c5frame,
               boxed title style={colback=c5frame, rounded corners}},
}

\definecolor{decocolor}{RGB}{255, 243, 224}   
\definecolor{scaffcolor}{RGB}{224, 240, 255}  
\definecolor{origcolor}{RGB}{235, 255, 235}   
 
\tcbset{
  perturbbox/.style={
    breakable,
    enhanced,
    boxrule=0.4pt,
    arc=3pt,
    left=6pt, right=6pt, top=4pt, bottom=4pt,
    fontupper=\small,
  }
}

\usepackage{microtype}

\usepackage{inconsolata}

\usepackage{graphicx}

\title{Beyond Accuracy: ARIA-Rubrics for Evaluating Audio Reasoning in Large Audio Language Models}

\author{
    Yupei Li\orcidlink{0009-0003-0832-1984} \\ Imperial College London, UK \\ \texttt{yl7622@ic.ac.uk} \\
    \And
    Qiyang Sun\orcidlink{0009-0001-9228-4543} \\ Imperial College London, UK \\ \texttt{q.sun23@imperial.ac.uk} \\
    \AND
   \hspace{-1.5cm} Mohamed Mady\orcidlink{0009-0007-8599-2174} \\ \hspace{-1.5cm}Technische Universität München\\ \hspace{-1.5cm}München, Germany \\ \hspace{-1.5cm}\texttt{mohamed.mady@tum.de} \\
    \And
    \hspace{-0.3cm}Chenxi Wang\orcidlink{0009-0006-3611-0000} \\ \hspace{-0.3cm}Mohamed bin Zayed University of Artificial Intelligence\\ \hspace{-0.3cm}Abu Dhabi, AE \\ \hspace{-0.3cm}\texttt{chenxi.wang@mbzuai.ac.ae} \\
    \AND
    Zhengwei Gong\orcidlink{0009-0006-5839-8252} \\ Shanghai Jiao Tong University, Shanghai, CN \\ \texttt{gzwsjtu98@sjtu.edu.cn} \\
    \And
    \hspace{0.8cm}Berrak Sisman\orcidlink{0000-0001-8078-3305} \\ \hspace{0.8cm}Johns Hopkins University, Maryland, US \\ \hspace{0.8cm}\texttt{sisman@jhu.edu} \\
        \AND
    Bj\"orn Schuller\orcidlink{0000-0002-6478-8699} \\ Technische Universität München, München, Germnay \\ \texttt{schuller@tum.de} \\
}

\begin{document}

\maketitle

\begin{abstract}

Large Audio Language Models (LALMs) have shown strong performance on audio reasoning benchmarks, but accuracy alone cannot distinguish true reasoning from superficial pattern matching, often overestimating reasoning ability since high scores may result from guessing rather than genuine audio understanding. Evaluating the reasoning process itself is essential for improving LALMs' reasoning ability, yet remains challenging. Existing methods either rely on costly human annotation or opaque LLM-as-judge approaches, making them impractical, biased, and lacking transparency. Moreover, audio reasoning introduces unique challenges absent in text-based settings, perceptual hallucination and cross-modal alignment between audio understanding and textual inference, hence text-based evaluation frameworks cannot be directly applied. Therefore, we propose ARIA-Rubrics (Audio Reasoning Integrity Assessment), a lightweight, annotation-free gold reasoning chains, automatic and transparent framework comprising six complementary metrics that evaluate audio reasoning quality across perceptual grounding, reasoning coherence, and answer consistency. We use Chain-of-Thought prompting as an externalization mechanism to make the reasoning process observable. Experiments on 9 models across 2 benchmarks identify three reasoning modes of current LALMs with actionable directions for future development, with ARIA-Rubrics achieving high correlation with 
human judgments. The code is available at the
\href{https://github.com/glam-imperial/allm_assessment}{GitHub Repository}
\end{abstract}

\section{Introduction}
Reasoning is a fundamental yet challenging capability for large language models (LLMs) \citep{plaat2025multi}, and becomes even more demanding for large audio language models (LALMs), where additional acoustic modalities need to be perceived, aligned, and integrated into the thinking process \citep{wang2024exploringreasoningabilitiesmultimodal}. Several benchmark datasets have been developed to assess LALMs' reasoning ability, such as MMAR \citep{ma2026mmar} and MMAU \citep{sakshi2025mmau}. 
To tackle these challenging benchmarks, a range of models have been developed using diverse pre-training backbones 
combined with post-training strategies including supervised fine-tuning \citep{das2024speechverse}, reinforcement learning \citep{wu2026audio}, and chain-of-thought (CoT) training \citep{ma2025audio}. While existing models reviewed in Section \ref{sec:related} have shown great potential \citep{su2025audio}, current evaluations rely primarily on accuracy or task completion rates, which provide a biased view of reasoning capabilities and cannot reveal whether models genuinely reason or merely exploit statistical patterns, a distinction crucial for understanding their limitations.

Therefore, assessing the reasoning process itself is essential, as it can uncover critical failure modes \citep{shojaee2026illusion} and guide future model improvements: \textit{Reasoning}, where the CoT faithfully drives the derivation of the correct answer; \textit{Scaffolding}, where the reasoning appears structurally complete but is vacuous or irrational, failing to meaningfully support the correct answer; and \textit{Decoration}, where the reasoning is entirely inconsistent and has no causal effect on the final answer. Existing approaches to process-level evaluation reviewed in Section \ref{sec:related} fall into two categories: human annotation \citep{yang2025speechr, huang2025step}, which is costly and labour-intensive at scale, and LLM-as-judge methods \citep{bi2026judgeboard}, which suffer from opacity and limited reliability. 

Textual LLMs have inspired evaluation methods beyond human annotation and LLM judges. For example, \citet{golovneva2023roscoesuitemetricsscoring} proposed multi-dimensional metrics for semantic consistency and coherence, while \citet{prasad2023receval} measured the informativeness of each reasoning step. However, these approaches do not directly apply to audio reasoning, which requires perceptual grounding: the quality of acoustic perception underlies all subsequent reasoning, and fabricated or inaccurate observations cannot be detected by text-only metrics. These limitations motivate the design of audio-native evaluation dimensions.

Building on the aforementioned limitations, we propose ARIA-Rubrics (Audio Reasoning Integrity Assessment) to systematically examine the intermediate reasoning process. Using CoT to reveal how models reach their answers, we design six complementary metrics covering acoustic perception, reasoning step progressiveness, and conclusion consistency. A combination of automatic metrics, lightweight model scoring, and targeted LLM judgments balances transparency and effectiveness, reducing dependence on opaque black-box judgments while leveraging natural language understanding where needed. Experiments on 9 models across MMAR and MMAU-mini reveal the three reasoning modes described above, providing diagnostic insights and actionable guidance for future development, with reliability confirmed through human evaluation.

In summary, our \textbf{contributions} are as follows. First, we propose \textbf{annotation-free gold reasoning chains, transparent, and automatic ARIA-Rubrics}, the first audio reasoning quality evaluation framework to the best of our knowledge, comprising six carefully designed complementary metrics covering both audio-native perceptual grounding and reasoning process quality. Second, we conduct comprehensive experiments on 9 models across 2 benchmarks, identifying three reasoning modes of current LALMs, \textit{Reasoning}, \textit{Scaffolding}, and \textit{Decoration}, and providing actionable directions for future model development. Third, we validate ARIA-Rubrics through human evaluation, demonstrating strong correlation with human judgments and confirming the reliability of our automated metrics.

\section{Related work}
\label{sec:related}
LALMs have rapidly advanced with diverse architectures and training paradigms in audio reasoning. Pretrained models such as Qwen2-Audio \citep{yang2024qwen2technicalreport}, Qwen2.5-Omni \citep{xu2025qwen25omnitechnicalreport}, Phi-4 Multimodal \citep{abouelenin2025phi}, and Gemma-3n \citep{gemmateam2025gemma3technicalreport} provide strong baselines through large-scale audio-text pretraining, while post-training strategies including reinforcement learning (R1-AQA \citep{wu2026audio}) and structured CoT fine-tuning (Audio-Reasoner \citep{ma2025audio}, AudSemThinker \citep{wijngaard2026audsemthinker}), further boost performance. Closed-source models such as GPT-4o Audio\footnote{\url{https://developers.openai.com/api/docs/models/gpt-4o-audio-preview}} and Gemini-2.5 Flash\citep{comanici2025gemini25pushingfrontier} represent the commercial frontier. These models are evaluated on benchmarks including MMAU \citep{sakshi2025mmau}, MMAR \citep{ma2026mmar}, AudioBench \citep{wang2025audiobench}, MUSE \citep{carone2026muse}, and AIR-Bench \citep{yang2024airbench}, covering diverse audio understanding and reasoning tasks. However, since these benchmarks rely solely on accuracy or task completion, they cannot distinguish genuine reasoning from statistical pattern exploitation. They reflect answer correctness, but not the quality or faithfulness of the reasoning process.

Existing frameworks for audio reasoning evaluation include MMAR-Rubrics \citep{ma2026interspeech}, which relies on proprietary models and human-annotated gold CoTs, limiting reproducibility and scalability; CAFE \citep{mao2026scalingfailsmitigatingaudio}, which uses LLM-as-judge to quantify perception and utilization errors but remains a black-box approach; and OCRA \citep{sedlavcek2025orca}, which depends entirely on human annotators without detailed guidance. In contrast, ARIA-Rubrics offers a lightweight, transparent, annotation-free gold reasoning chains framework with audio-native metrics, requiring neither large LLMs nor human annotations.

Evaluating reasoning chain quality has been widely studied in NLP. CoT prompting \citep{wei2022chain} externalizes intermediate reasoning steps, making them observable and assessable, and their quality has been shown to correlate strongly with final answer correctness \citep{lightman2024lets}, establishing CoT as a reliable proxy for reasoning ability. Accordingly, many evaluation rubrics leverage CoT. For instance, ROSCOE \citep{golovneva2023roscoesuitemetricsscoring} measures semantic consistency and coherence between steps, ReCEval \citep{prasad2023receval} assesses correctness and informativeness via natural language inference, Process Reward Models \citep{zheng2025process} verify step-level mathematical reasoning using human-annotated supervision, FaithScore \citep{jing2024faithscore} evaluates faithfulness to source context through atomic fact verification, and SocREval \citep{he2024socreval} applies the Socratic method for annotation-free gold reasoning chains reasoning assessment. 
These automatic metrics avoid costly human annotation. However, these frameworks are designed for text-only settings and lack perceptual grounding dimensions, and perceptual hallucinations cannot be detected by text-based metrics, but the aspects motivate the design of audio-native evaluation dimensions in ARIA-Rubrics.

\section{ARIA-Rubrics framework}
\subsection{Overview}
\begin{figure*}[ht]
    \centering
    \includegraphics[width=0.89\textwidth]{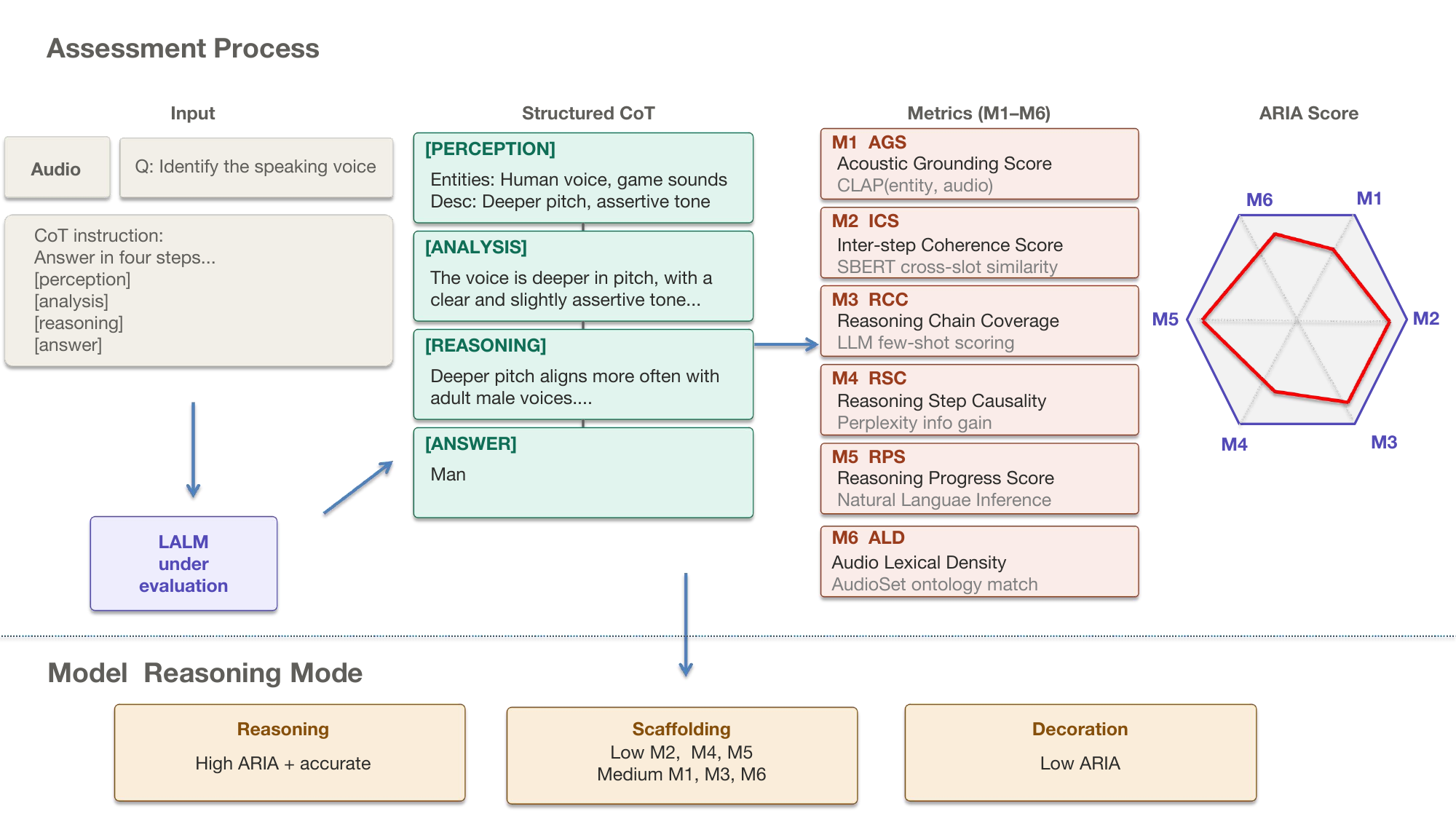}
    \caption{Overview of the ARIA-Rubrics evaluation pipeline. Given an audio input and a question, the candidate LALM is prompted with a structured four-step CoT template to externalize its reasoning process. The resulting CoT output is evaluated by six complementary metrics (M1--M6) covering hallucination detection, cross-slot coherence, content substantiveness, reasoning progress, reasoning--answer consistency, and audio vocabulary density. A final ARIA score is computed as the mean of all six metrics. Based on the score profile, each model could be categorised into one of three reasoning modes: \textit{Reasoning}, \textit{Scaffolding}, or \textit{Decoration}.}
    \vspace{-0.5cm}
    \label{fig:pipeline}
\end{figure*}

ARIA-Rubrics is designed to evaluate the reasoning quality of LALMs by examining the intermediate reasoning process rather than the final answer alone. The core insight is that a faithful reasoning process should be grounded in accurate acoustic perception, progress coherently toward the answer, and maintain consistency between reasoning and conclusion. To make the reasoning process observable, we adopt CoT prompting as an externalization mechanism, eliciting models to produce structured reasoning chains rather than direct answers. Given this structured output, ARIA-Rubrics evaluates six complementary metrics (M1--M6) targeting different dimensions of reasoning quality. The overall pipeline is illustrated in Figure~\ref{fig:pipeline}. The final ARIA score is computed as the mean of all six metrics. 

\subsection{CoT prompt design}
As previously stated, we design a structured four-step CoT prompt comprising \textit{Perception}, \textit{Analysis}, \textit{Reasoning}, and \textit{Answer}. Each step serves a distinct cognitive function: \textit{Perception} requires the model to enumerate the sound entities it actually hears and describe their acoustic properties and temporal order without interpretation; \textit{Analysis} connects acoustic observations to their possible meanings without drawing final conclusions; \textit{Reasoning} weighs the evidence and eliminates alternatives to arrive at a conclusion; and \textit{Answer} states the final choice, with the actual prompt shown in Appendix \ref{app:prompt_CoT}.

We design the four-step structure by examining LLM-generated CoT annotations from ALR \citep{diao2025soundmind} and CoTA \citep{xie2025audioreasonerimprovingreasoningcapability}, identifying a consistent cognitive flow from acoustic perception to semantic inference as their core reasoning pattern, and human-annotated CoT examples remain unavailable in open-source form (e.g., MMAR \citep{ma2026mmar}). This design choice is further supported by \citet{ma2025audio}, 
who
showed that different CoT template variants, including zero-shot, manual few-shot, and other prompt formats, yield accuracy differences within 3\% absolute, suggesting that prompt engineering alone does not significantly constrain inherent reasoning capabilities in a biased assessment framework. We further validate our specific prompt design through a sanity check comparing model accuracy with and without our CoT prompt across all evaluated models, reported in Appendix \ref{app:cot_vs_baseline}. Results show no huge performance difference in most cases, confirming that our structured prompt does not restrict model reasoning ability, and that observed accuracy degradation under CoT prompting more likely reflects the model's lack of genuine reasoning ability rather than prompt interference.

\subsection{Six Complementary Metrics}
\subsubsection{Acoustic Grounding Score (M1)}

Since perception is a fundamental component of audio reasoning, we introduce the first metric specifically designed to evaluate acoustic perception ability. The Acoustic Grounding Score (AGS) measures whether the sound entities identified during the \textit{Perception} stage are genuinely grounded in the input audio signal. This metric detects perceptual hallucinations, where models fabricate nonexistent acoustic events or misidentify entities. For each entity 
$e_i$ listed in the \textit{Entities} line of \textit{Perception}, we compute its similarity to the whole audio signal using a 
Contrastive Language-Audio Pretraining (CLAP) model \citep{laionclap2023}, a pre-trained model to align audio and text representations in a shared embedding space. The AGS is defined as:
\begin{equation}
\text{AGS} = \frac{1}{|E|} \sum_{i=1}^{|E|} \text{CLAP\_sim}(e_i, \mathbf{a}),
\end{equation}
where $E$ is the set of extracted entities and $\mathbf{a}$ is the whole audio input. It also prevents any hack strategies in which models output as many sound entities as possible to artificially increase the average score. If no entities are detected in \textit{Perception}, M1 is 0.

\subsubsection{Inter-step Coherence Score (M2)}
After the perception stage is recognized, a reliable reasoning model should correctly utilize the extracted evidence for subsequent reasoning. The Inter-step Coherence Score (ICS) evaluates semantic coherence between consecutive reasoning steps, assessing whether each step is grounded by prior context. Specifically, coherence is computed across the first three reasoning steps (excluding the final answer, which is a single output) using sentence-level embeddings \citep{song2020mpnet} (\textit{all-mpnet-base-v2}), chosen for its strong performance on downstream text understanding and support for long-context embeddings. The ICS is defined as: 

\begin{equation}
\text{ICS} = \tfrac{1}{2}\bigl(c(f(A),f(P))+c(f(R),f(A))\bigr)
\end{equation}
where $f(\cdot)$ denotes the sentence embedding function, $c(\cdot)$ represents cosine similarity, $P,A,R$ correspond to the sentences from the \textit{Perception}, \textit{Analysis}, and  \textit{Reasoning} sections, respectively. If either slot is absent, M2 is 0. 

\subsubsection{Reasoning Chain Coverage (M3)}
Though the references from previous steps can be verified, we also need to assess whether each step in the CoT contains substantive content rather than meaningless contents. Accordingly, the Reasoning Chain Coverage (RCC) is computed as follows: for each slot $s \in \{\textit{Perception}, \textit{Analysis}, \textit{Reasoning}\}$, we employ Qwen2.5-1.5B-Instruct as a lightweight scorer with few-shot prompting to assign a substantiveness score $\sigma(s) \in [0, 1]$, where 0 corresponds to entirely empty or repetitive content, and 1 corresponds to highly specific and informative content. The prompt for the scorer is shown in Appendix \ref{app:RCC}. The \textit{Answer} slot is scored deterministically: 1.0 if it is present and non-empty, and 0.0 otherwise. 
The RCC is computed as the average of four slot scores. If a slot is absent, its corresponding score $(\sigma(P/A/R/Ans)$, where $Ans$ denotes sentences from the \textit{Answer} section) is set to 0.

We adopt an LLM-as-a-judge strategy, noting that the task is relatively simple and mainly requires basic language understanding. Few-shot examples in the evaluation prompts improve reliability, proved by \citet{zhang2023prompt}. Unlike fully black-box LLM-based evaluations of overall reasoning ability, our framework leverages LLMs’ strengths in language understanding while avoiding tasks where their judgments are unreliable.

\subsubsection{Reasoning Step Causality (M4)}

The reasoning process should be complete and coherent, and also progressively guide toward the correct answer. Reasoning Step Causality (RSC) measures whether each step advances toward the ground-truth answer by estimating the information gain contributed by that step. Specifically, we track how the probability of the correct answer changes as each step is added to the context, using a small language model to compute conditional perplexity. If an added step makes the correct answer more predictable, the perplexity decreases; we define information gain as the reduction in perplexity relative to the previous context. To ensure a non-negative score, we apply a sigmoid transformation to the information gain, yielding the exponential form below. This design follows ROSCOE \citep{golovneva2023roscoesuitemetricsscoring}, which assesses step-level informativeness without human-annotated references. We use Qwen2.5-1.5B-Instruct (as in M3) as the evaluator to avoid reliance on additional pretrained models.


\newcommand{\ppl}{\mathrm{PPL}}

\begin{equation}
\begin{aligned}
\text{Info}_A &= \ppl(\hat{Ans} \mid P, A) - \ppl(\hat{Ans} \mid P) \\
\text{Info}_R &= \ppl(\hat{Ans} \mid P, A, R) \\
              &\quad - \ppl(\hat{Ans} \mid P, A) \\
\text{RSC}  &= \frac{1}{2} \sum_{i \in \{A,R\}} \frac{1}{1 + e^{\text{Info}_i}}
\end{aligned}
\end{equation}
where $\hat{Ans}$ is the groundtruth Answer. 
If a slot is absent, the corresponding score ($\text{Info}_{A/R}$) is 0.

\subsubsection{Reasoning Progress Score (M5)}
Collecting correct information is insufficient if the model does not progress fluently. The Reasoning Progress Score (RPS) evaluates whether each step is conditioned on the previous one, distinguishing it from M2, which measures whether prior context is referenced. We adopt Natural Language Inference (NLI), a well-studied task for determining whether a premise entails a hypothesis, as a natural fit for this metric. We use Roberta-large-mnli \citep{liu2019robertarobustlyoptimizedbert} fine-tuned on NLI tasks as the classifier. The RPS is calculated as:
\begin{equation}
    \text{RPS} = \frac{1}{3} \bigl( g(P,A) 
+ g(A,R)
 + g(R,Ans)  \bigr),
\end{equation}

where $g(\cdot)$ is the NLI model’s entailment score. If a slot is absent, its corresponding score is 0. 

\subsubsection{Audio Lexical Density (M6)}

After evaluating all reasoning processes, a common criticism of LLMs is that their reasoning chains are often overly long \citep{an2026don}, potentially due to mechanisms such as self-consistency checks \citep{wang2022selfconsistency} or exploration of multiple solution paths \citep{wang2025don}. However, excessive length may cause models to lose focus or produce unnecessarily verbose reasoning. We propose Audio Lexical Density (ALD), which measures the proportion of audio-relevant vocabulary in the reasoning chain. We hypothesize that, in audio reasoning tasks, greater use of audio-related terms correlates with more meaningful inference. Although this metric is an approximation, it can effectively penalize excessive self-reflection without progress and the inclusion of irrelevant content.

Specifically, we utilize the AudioSet ontology \citep{gemmeke2017audio}, which contains a comprehensive set of audio-related terms, including verbal actions, sound-producing subjects, and environmental sounds. We deem these are complete and enough as these are annotated by human experts. The audio reasoning benchmarks mainly concern these domains. We manually filtered out some synonymous terms (e.g., baby and infant), leaving a total of 417 words (listed in Appendix \ref{app:ontology}). For each word in the CoT output, we compute its similarity to each ontology term using the word2vec model\footnote{\url{https://huggingface.co/fse/word2vec-google-news-300}} \citep{mikolov2013efficientestimationwordrepresentations}, selecting the maximum similarity as the score for that word. The overall ALD score is then obtained by averaging these maximum similarities across all words in the reasoning chain.

\subsection{Overall discussion}
The six metrics are designed to be complementary rather than redundant, each targeting a distinct dimension of reasoning quality while collectively preventing gaming through mutual constraints. Specifically, artificially inflating any single metric will inevitably suppress others: a model that fabricates plausible-sounding entities to boost M6 will be penalised by M3 if those entities are repetitively listed; a model that rushes to the correct answer too quickly to maximise M5 will exhibit low M1 and M4 due to insufficient perceptual grounding; and a model recalling too much to increase M2 will lead to decrease in M5. Together, the six metrics cover the full reasoning pipeline from acoustic perception to final answer, and the ARIA score is computed as their unweighted mean, yielding a value $\in [0,1]$.

The metric profile also provides a diagnostic basis for identifying the three reasoning modes. Models exhibiting high accuracy but low ARIA scores are characterised as \textit{Decoration}. Conversely, high ARIA with high accuracy indicates genuine \textit{Reasoning}. The \textit{Scaffolding} mode is primarily identified through low scores on M2, M4, and M5, which assess the semantic content and meaningfulness of the reasoning process; a model may maintain structural coherence (high M3) while producing vacuous or disconnected content that fails these meaning-oriented checks.

\section{Experiments and Results}
We evaluate ARIA-Rubrics on two audio reasoning benchmarks: MMAU-mini \citep{sakshi2025mmau} and MMAR \citep{ma2026mmar}, each comprising 1,000 multiple-choice questions with no sample overlap between the two datasets. To validate the effects of our rubrics, we select 9 LALMs spanning diverse base models and training paradigms, covering both open-source and closed-source systems. Detailed hyperparameter settings and implementation details are provided the Appendix \ref{app:hyper}. Results are presented in Table \ref{tab:main_results}.
We extract final answers from model outputs via a three-stage matching procedure: exact string matching against answer choices, substring matching, and letter-to-choice index mapping. The full details are provided in Appendix~\ref{app:extraction}.

\newcommand{\rc}[2]{\cellcolor{warm#1}{#2}}

\begin{table*}[t]
\centering
\caption{Results on MMAR and MMAU-mini benchmarks. Cell color indicates per-column rank (darker = higher rank, 9-level warm gradient). ARIA = mean of M1--M6. Acc means accuracy.}
\label{tab:main_results}
\renewcommand{\arraystretch}{1.2}
\resizebox{0.9\textwidth}{!}{%
\begin{tabular}{l|cc|cc|cc|cc|cc|cc|c|c}
\toprule
\textbf{Model}
  & \multicolumn{2}{c|}{\textbf{M1}}
  & \multicolumn{2}{c|}{\textbf{M2}}
  & \multicolumn{2}{c|}{\textbf{M3}}
  & \multicolumn{2}{c|}{\textbf{M4}}
  & \multicolumn{2}{c|}{\textbf{M5}}
  & \multicolumn{2}{c|}{\textbf{M6}}
  & \textbf{ARIA}
  & \textbf{Acc(\%)} \\
\midrule
\multicolumn{15}{c}{\textit{MMAR}} \\
\midrule
\multicolumn{15}{l}{\quad\textit{Open-source Models}} \\
\midrule
Qwen2-Audio
  & \rc{9}{0.1963} & \rc{9}{±0.1988}
  & \rc{7}{0.6169} & \rc{7}{±0.1917}
  & \rc{7}{0.6967} & \rc{7}{±0.1711}
  & \rc{8}{0.5220} & \rc{8}{±0.2757}
  & \rc{8}{0.2015} & \rc{8}{±0.1792}
  & \rc{8}{0.4326} & \rc{8}{±0.0815}
  & \rc{8}{0.4443} & \rc{9}{17.7\%} \\
Qwen2.5-Omni
  & \rc{1}{0.4344} & \rc{1}{±0.1722}
  & \rc{3}{0.6815} & \rc{3}{±0.2202}
  & \rc{3}{0.7561} & \rc{3}{±0.1987}
  & \rc{5}{0.5887} & \rc{5}{±0.2734}
  & \rc{7}{0.2873} & \rc{7}{±0.2160}
  & \rc{3}{0.5215} & \rc{3}{±0.1238}
  & \rc{3}{0.5449} & \rc{4}{44.3\%} \\
Gemma-3n-E2B
  & \rc{6}{0.2861} & \rc{6}{±0.1754}
  & \rc{4}{0.6442} & \rc{4}{±0.1972}
  & \rc{4}{0.7362} & \rc{4}{±0.1720}
  & \rc{3}{0.6315} & \rc{3}{±0.2934}
  & \rc{5}{0.3215} & \rc{5}{±0.2140}
  & \rc{7}{0.4356} & \rc{7}{±0.0944}
  & \rc{4}{0.5092} & \rc{6}{27.3\%} \\
R1-AQA
  & \rc{8}{0.2107} & \rc{8}{±0.1900}
  & \rc{9}{0.5700} & \rc{9}{±0.2004}
  & \rc{9}{0.6891} & \rc{9}{±0.1589}
  & \rc{7}{0.5358} & \rc{7}{±0.2764}
  & \rc{9}{0.1812} & \rc{9}{±0.1699}
  & \rc{9}{0.4237} & \rc{9}{±0.0779}
  & \rc{9}{0.4351} & \rc{8}{21.6\%} \\
Audio-Reasoner
  & \rc{7}{0.2401} & \rc{7}{±0.1747}
  & \rc{8}{0.6159} & \rc{8}{±0.2584}
  & \rc{8}{0.6293} & \rc{8}{±0.2369}
  & \rc{4}{0.6099} & \rc{4}{±0.2191}
  & \rc{6}{0.2887} & \rc{6}{±0.2344}
  & \rc{5}{0.4817} & \rc{5}{±0.1202}
  & \rc{6}{0.4776} & \rc{7}{20.1\%} \\
AudSemThinker
  & \rc{5}{0.3309} & \rc{5}{±0.1923}
  & \rc{6}{0.4626} & \rc{6}{±0.3480}
  & \rc{6}{0.5735} & \rc{6}{±0.3408}
  & \rc{6}{0.5672} & \rc{6}{±0.2366}
  & \rc{4}{0.2317} & \rc{4}{±0.2313} 
  & \rc{1}{0.6171} & \rc{1}{±0.2106}
  & \rc{7}{0.4638} & \rc{5}{31.1\%} \\
Phi-4
  & \rc{4}{0.3432} & \rc{4}{±0.1665}
  & \rc{5}{0.5972} & \rc{5}{±0.2820}
  & \rc{5}{0.6481} & \rc{5}{±0.2538}
  & \rc{9}{0.5129} & \rc{9}{±0.2499}
  & \rc{3}{0.2210} & \rc{3}{±0.2029}
  & \rc{2}{0.6143} & \rc{2}{±0.1497}
  & \rc{5}{0.4895} & \rc{3}{27.1\%} \\
\midrule
\multicolumn{15}{l}{\quad\textit{Closed-source Models}} \\
\midrule
GPT-4o-audio
  & \rc{3}{0.3725} & \rc{3}{±0.1686}
  & \rc{1}{0.7142} & \rc{1}{±0.0849}
  & \rc{1}{0.7838} & \rc{1}{±0.0589}
  & \rc{1}{0.7506} & \rc{1}{±0.2522}
  & \rc{1}{0.4346} & \rc{1}{±0.2057}
  & \rc{6}{0.4219} & \rc{6}{±0.0608}
  & \rc{1}{0.5796} & \rc{1}{67.6\%} \\
Gemini 2.5 Flash
  & \rc{2}{0.3808} & \rc{2}{±0.1630}
  & \rc{2}{0.6776} & \rc{2}{±0.0908}
  & \rc{2}{0.7729} & \rc{2}{±0.0730}
  & \rc{2}{0.7426} & \rc{2}{±0.2379}
  & \rc{2}{0.4021} & \rc{2}{±0.1938}
  & \rc{4}{0.4375} & \rc{4}{±0.0557}
  & \rc{2}{0.5689} & \rc{2}{62.6\%} \\
\midrule
\multicolumn{15}{c}{\textit{MMAU-mini}} \\
\midrule
\multicolumn{15}{l}{\quad\textit{Open-source Models}} \\
\midrule
Qwen2-Audio
  & \rc{8}{0.1941} & \rc{8}{±0.1315}
  & \rc{8}{0.6123} & \rc{8}{±0.2075}
  & \rc{8}{0.6959} & \rc{8}{±0.1871}
  & \rc{9}{0.5346} & \rc{9}{±0.2180}
  & \rc{9}{0.2209} & \rc{9}{±0.1873}
  & \rc{8}{0.4462} & \rc{8}{±0.0559}
  & \rc{9}{0.4507} & \rc{9}{24.1\%} \\
Qwen2.5-Omni
  & \rc{9}{0.1929} & \rc{9}{±0.1305}
  & \rc{3}{0.6654} & \rc{3}{±0.1958}
  & \rc{2}{0.7587} & \rc{2}{±0.1681}
  & \rc{5}{0.6364} & \rc{5}{±0.2624}
  & \rc{3}{0.3481} & \rc{3}{±0.1903}
  & \rc{3}{0.4771} & \rc{3}{±0.0705}
  & \rc{5}{0.5131} & \rc{2}{58.4\%} \\
Gemma-3n-E2B
  & \rc{5}{0.3174} & \rc{5}{±0.1764}
  & \rc{6}{0.6207} & \rc{6}{±0.1908}
  & \rc{6}{0.7181} & \rc{6}{±0.1655}
  & \rc{3}{0.6903} & \rc{3}{±0.2784}
  & \rc{4}{0.3237} & \rc{4}{±0.2064}
  & \rc{6}{0.4491} & \rc{6}{±0.0943}
  & \rc{4}{0.5199} & \rc{6}{36.5\%} \\
R1-AQA
  & \rc{7}{0.2670} & \rc{7}{±0.1815}
  & \rc{9}{0.5847} & \rc{9}{±0.2072}
  & \rc{9}{0.6926} & \rc{9}{±0.1737}
  & \rc{7}{0.5536} & \rc{7}{±0.2775}
  & \rc{8}{0.2056} & \rc{8}{±0.1653}
  & \rc{9}{0.4490} & \rc{9}{±0.0898}
  & \rc{8}{0.4588} & \rc{8}{26.3\%} \\
Audio-Reasoner
  & \rc{6}{0.3032} & \rc{6}{±0.1636}
  & \rc{7}{0.5774} & \rc{7}{±0.2946}
  & \rc{7}{0.6186} & \rc{7}{±0.2646}
  & \rc{4}{0.6877} & \rc{4}{±0.2029}
  & \rc{6}{0.2860} & \rc{6}{±0.2334}
  & \rc{2}{0.5097} & \rc{2}{±0.1622}
  & \rc{6}{0.4971} & \rc{7}{34.0\%} \\
AudSemThinker
  & \rc{3}{0.3637} & \rc{3}{±0.1764}
  & \rc{5}{0.5237} & \rc{5}{±0.3325}
  & \rc{5}{0.6178} & \rc{5}{±0.3120}
  & \rc{8}{0.5476} & \rc{8}{±0.2286}
  & \rc{7}{0.2616} & \rc{7}{±0.2239}
  & \rc{1}{0.6046} & \rc{1}{±0.1973}
  & \rc{7}{0.4865} & \rc{3}{46.0\%} \\
Phi-4
  & \rc{4}{0.3443} & \rc{4}{±0.1724}
  & \rc{4}{0.6507} & \rc{4}{±0.2004}
  & \rc{4}{0.7300} & \rc{4}{±0.1650}
  & \rc{8}{0.6255} & \rc{8}{±0.2730}
  & \rc{5}{0.2744} & \rc{5}{±0.1918}
  & \rc{4}{0.5521} & \rc{4}{±0.1090}
  & \rc{3}{0.5295} & \rc{4}{40.7\%} \\
\midrule
\multicolumn{15}{l}{\quad\textit{Closed-source Models}} \\
\midrule
GPT-4o-audio
  & \rc{2}{0.4035} & \rc{2}{±0.1632}
  & \rc{1}{0.6836} & \rc{1}{±0.1194}
  & \rc{1}{0.7780} & \rc{1}{±0.0969}
  & \rc{2}{0.7491} & \rc{2}{±0.2392}
  & \rc{1}{0.4161} & \rc{1}{±0.1975}
  & \rc{7}{0.4502} & \rc{7}{±0.0752}
  & \rc{2}{0.5801} & \rc{1}{55.8\%} \\
Gemini 2.5 Flash
  & \rc{1}{0.4199} & \rc{1}{±0.1472}
  & \rc{2}{0.6564} & \rc{2}{±0.1025}
  & \rc{3}{0.7680} & \rc{3}{±0.0805}
  & \rc{1}{0.7763} & \rc{1}{±0.2253}
  & \rc{2}{0.4169} & \rc{2}{±0.1818}
  & \rc{5}{0.4541} & \rc{5}{±0.0704}
  & \rc{1}{0.5819} & \rc{5}{51.8\%} \\
\bottomrule
\end{tabular}}
\end{table*}

We observe \textbf{a systematic divergence between ARIA scores and accuracy rankings}, indicating that accuracy alone provides an incomplete picture of reasoning quality. Two representative cases illustrate this (Table \ref{tab:case_study}): First, a model may achieve high accuracy despite low ARIA scores; for example, AudSemThinker (M3=0.36, M5=0.50) answers correctly, but its CoT lacks acoustic grounding: the Perception step only assigns an accent label without phonetic evidence, and the Reasoning step circularly restates this label without inferential progression. Second, a model may score high on ARIA despite an incorrect answer; Qwen2.5-Omni (M3=0.84, M5=1) demonstrates genuine acoustic observation and logical reasoning, yet fails due to a knowledge gap about Indian English phonetics rather than reasoning failure. These cases show that ARIA provides diagnostic insights invisible to accuracy-based evaluation, enabling a faithful assessment of model reasoning ability, while focusing on reasoning rather than knowledge coverage. 

We additionally conducted an LLM-as-a-judge evaluation \cite{gu2026survey} to examine whether there is any divergence between ARIA scores and LLM-based judgements (Deepseek-v4-pro \cite{xu2026deepseek} and Gemini-v3-flash \cite{team2023gemini}), with prompts of LLM-as-a-judge in Appendix \ref{app:llm_as_a_judge}. The results in Table \ref{tab:aria_ranking_llm} show that the ranking produced by ARIA is largely consistent with the overall model performance assessed by the LLM judge, while providing more interpretable insights into specific directions for future improvement. Moreover, unlike LLM-based evaluation, which can be computationally and token intensive, ARIA provides a more efficient and structured evaluation framework.

\begin{table}[h]
\centering
\small
\caption{Comparison of overall accuracy (Acc), DeepSeek-based evaluation (DS), Gemini-based evaluation (Gem), and the resulting ARIA ranking.}
\resizebox{\columnwidth}{!}{
\begin{tabular}{lcccc}
\toprule
\textbf{Model} & \textbf{Acc} & \textbf{DS} & \textbf{Gem} & \textbf{ARIA Rank} \\
\midrule
GPT-4o-audio     & 0.676 & 0.776 & 0.786 & 1 \\
Gemini 2.5 Flash & 0.626 & 0.784 & 0.800 & 2 \\
Qwen2.5-Omni     & 0.443 & 0.497 & 0.488 & 3 \\
AudSemThinker    & 0.311 & 0.339 & 0.348 & 4 \\
Phi-4            & 0.254 & 0.250 & 0.233 & 5 \\
Gemma-3n-E2B     & 0.273 & 0.433 & 0.430 & 6 \\
Audio-Reasoner   & 0.201 & 0.263 & 0.271 & 7 \\
R1-AQA           & 0.216 & 0.288 & 0.256 & 8 \\
Qwen2-Audio      & 0.177 & 0.279 & 0.254 & 9 \\
\bottomrule
\end{tabular}
}
\label{tab:aria_ranking_llm}
\end{table}

\begin{table*}[ht]
\caption{Two illustrative cases from \textbf{ID:} 2Ey6mTjurYc\_00-00-00\_00-00-19 in MMAR highlight the divergence between ARIA scores and accuracy. AudSemThinker achieves high accuracy but low ARIA. Qwen2.5-Omni scores high ARIA despite an incorrect answer.}
\label{tab:case_study}
\centering
\small
\resizebox{0.9\textwidth}{!}{%

\begin{tabular}{lp{5.5cm}p{7.5cm}}
\toprule
\textbf{Models} & \textbf{AudSemThinker} & \textbf{Qwen2.5-Omni} \\
\midrule
\multicolumn{3}{l}{\textbf{Question:} Which region might the sixth pronunciation of ``bottle of water'' in the audio come from?} \\
\multicolumn{3}{l}{\textbf{Choices:} (A) India \quad (B) United Kingdom \quad (C) United States \quad (D) Australia} \\
\midrule
\textbf{CoT} 
& {[PERCEPTION] Entities: male voice, speech, male speech, narration, monologue, English, male, young adult, Indian accent. [ANALYSIS] The sixth pronunciation of ``bottle of water'' is characterized by an Indian accent. This suggests that the speaker \textbf{is likely from India or has an Indian accent}. The other options (United Kingdom, United States, Australia) are less likely based on the specific accent heard. [REASONING] The Indian accent is a strong indicator of the speaker's origin. The other options are less likely because they do not match the specific accent heard in the audio. [ANSWER] A. India}
& [PERCEPTION] Entities: speech, male voice, English.... Description: The audio features a male voice speaking in English... \textbf{The speech is slow and peaceful, and the vowels are consistent. The speech is clear and the pronunciation is stable throughout the audio.} [ANALYSIS] ... The repetition of the phrase suggests \textbf{a focus on the pronunciation} or perhaps a language learning exercise. The male voice is clear and the English is standard, without any regional accents. [REASONING] ... The lack of any regional accent suggests that the speaker \textbf{is likely from a region where English is spoken without a distinct regional accent}. ... Without a specific regional accent information, it is difficult to pinpoint the exact region. \textbf{However, the standard English pronunciation is more common in the United States and the United Kingdom. }[ANSWER] United States \\
\midrule
\textbf{Answer} & India \textcolor{green}{$\checkmark$} & United States \textcolor{red}{\texttimes} \\
\bottomrule
\end{tabular}
}
\vspace{-0.3cm}
\end{table*}

\paragraph{Model reasoning modes}
We further analyze model reasoning modes, providing additional validation of our rubrics. Specifically, we manually perturb CoTs on randomly selected instances, injecting two types of altered reasoning: a Decoration CoT, with unstructured and vacuous content (e.g., random words like 0\textit{I love dog}), and a Scaffolding CoT, with correct structure but fabricated content (e.g., \textit{the sound is alike bird and tiger}). We then combine the perturbed CoT with the question and audio to assess whether each model's final answer changes. Full prompts and responses are in Appendix~\ref{app:mm_output}, with results summarized in Table~\ref{tab:perturbation_modes}.
\begin{table}[h]
\centering
\small
\caption{Reasoning mode assignment based on CoT perturbation sanity check and ARIA score rank.}
\label{tab:perturbation_modes}
\resizebox{0.7\columnwidth}{!}{
\begin{tabular}{lll}
\toprule
\textbf{Mode} & \textbf{Model} & \textbf{ARIA Rank} \\
\midrule
\multirow{3}{*}{Reasoning}   & GPT-4o-audio     & 1 \\
                              & Gemini 2.5 Flash & 2\\
                              & Qwen2.5-Omni     & 3\\
\midrule
\multirow{3}{*}{Scaffolding} & Gemma-3n-E2B         & 4\\
                              & Phi-4            & 5\\
                              & Audio-Reasoner   & 6\\
\midrule
\multirow{3}{*}{Decoration}  & AudSemThinker    & 7\\
                              & R1-AQA           & 8\\
                             & Qwen2-Audio      & 9\\

\bottomrule
\end{tabular}}
        \vspace{-0.4cm}

\end{table}

The results show a strong alignment between perturbation-derived mode assignments and ARIA rankings: Reasoning models rank highest (1–3), Scaffolding models mid (5–7), and Decoration models lowest, confirming that ARIA reflects reasoning quality. Closer examination reveals distinct failure patterns: Scaffolding models, such as AudSemThinker, produce structurally complete but often irrational CoTs (e.g., ``low-pitched hooting → tiger''), occasionally reaching correct answers, consistent with low M2, M4, and M5 scores. Decoration models (R1-AQA, Qwen2-Audio) show no causal effect of CoT on the final answer. In contrast, Reasoning models resist misleading CoTs and re-ground their answers in the audio, demonstrating genuine cross-modal reasoning.

\paragraph{Dataset-independent analysis for ARIA rubrics}

\begin{figure}[h]
    \centering
    \includegraphics[width=0.9\linewidth]{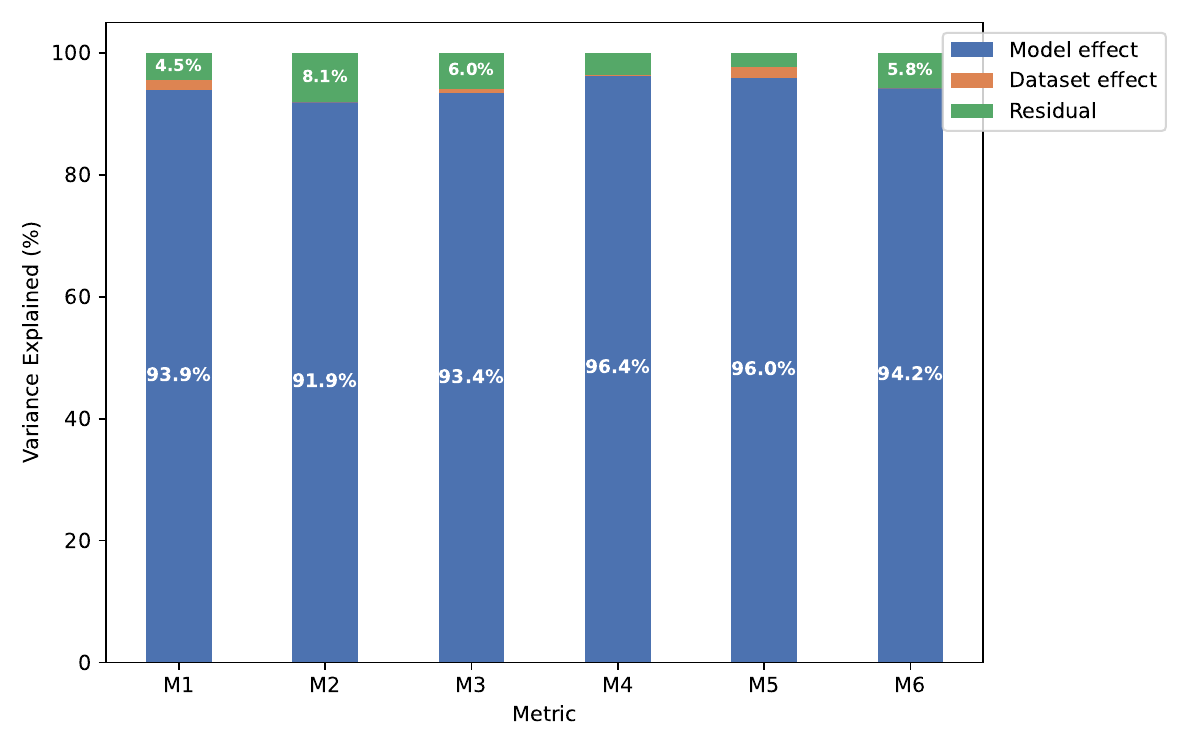}
    \caption{Two-way ANOVA decomposition of score variance across models and datasets. Blue bars indicate the proportion of variance explained by model identity ($\eta^2_{\text{model}}$), and orange bars indicate the proportion explained by dataset identity ($\eta^2_{\text{dataset}}$), for six metrics.}
    \vspace{-0.7cm}
    \label{fig:variance}
\end{figure}

\begin{figure}[h]
    \centering
    \includegraphics[width=0.9\linewidth]{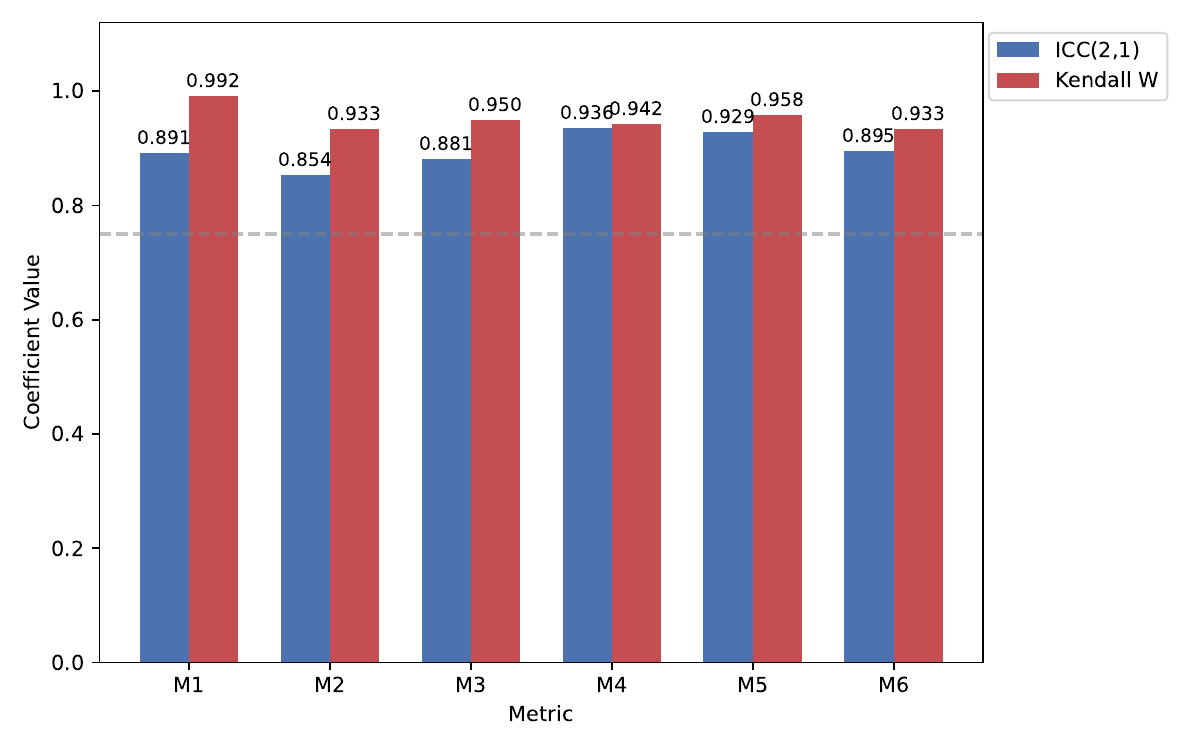}
    \caption{ICC(2,1) and Kendall's W coefficients for each metric, measuring cross-dataset consistency of model scores and rankings respectively.}
    \label{fig:icc}
        \vspace{-0.7cm}

\end{figure}

We apply ARIA rubrics across different reasoning datasets to demonstrate that they evaluate true LALM reasoning skills rather than dataset-specific question effects. The questions serve to externalize reasoning, akin to the CoT template. Three complementary analyses support this: a two-way Analysis of Variance (ANOVA) shows that model identity accounts for 91.9–96.4\% of score variance ($\eta^2_{\text{model}}$), while dataset identity contributes at most 1.7\%; Kendall’s $W \in [0.933, 0.992]$ indicates highly consistent model rankings across benchmarks; and ICC(2,1) confirms instance-level cross-dataset agreement (Figures \ref{fig:variance} and \ref{fig:icc}). These results demonstrate that ARIA captures reasoning capability independent of dataset, suggesting that a limited number of benchmarks suffices to reveal LALM reasoning ability.


\begin{figure}[h]
    \centering
    \includegraphics[width=0.8\linewidth]{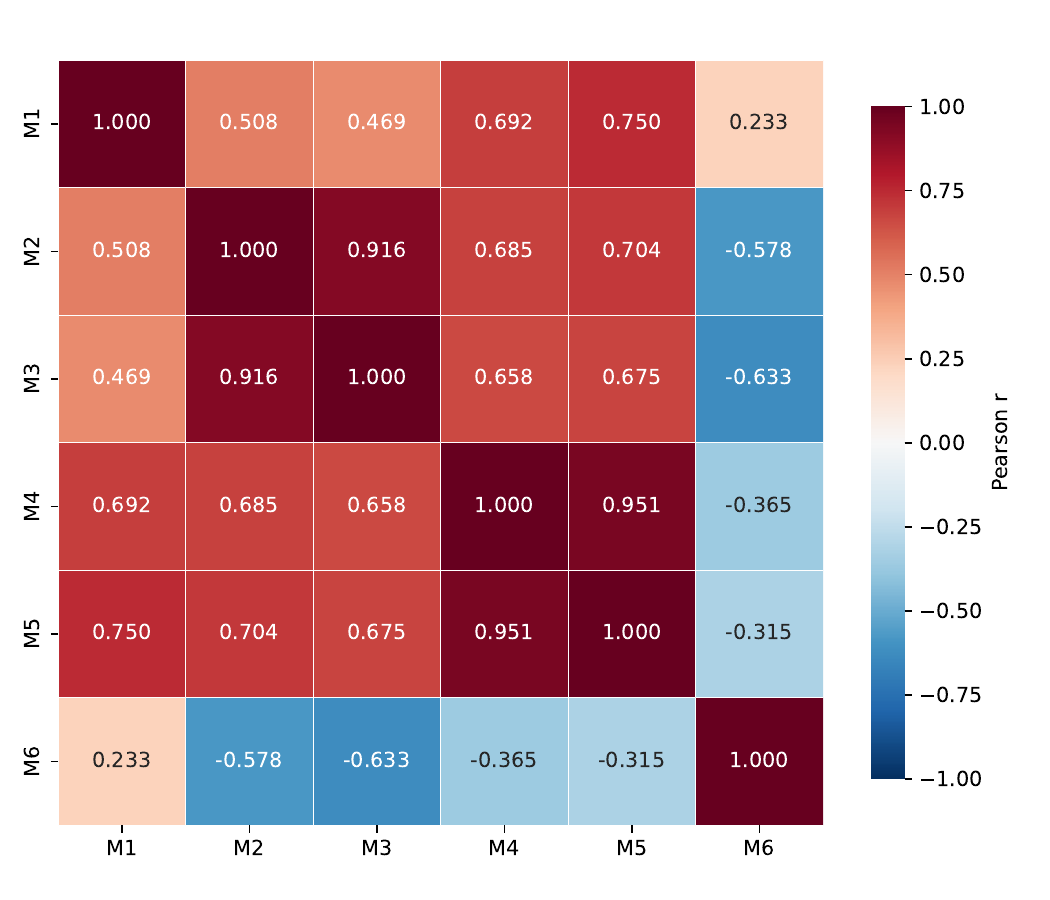}
    \caption{The Pearson correlation matrix across the six metrics.}
        \vspace{-0.7cm}

    \label{fig:pearson}
\end{figure}

\paragraph{Component's (M1-M6) metrics' correlation}
We compute correlations between ARIA components to assess their specific roles (Figure~\ref{fig:pearson}). M4 and M5 show the strongest correlation ($r = 0.951$), as expected as rare models gather causally relevant information but progress differently from the final answer. It is interesting that M2 and M3 are also highly correlated ($r = 0.916$), suggesting that models producing more substantive content per step tend to reference prior steps consistently, reflecting iterative reinforcement of coherence.

The remaining metric pairs show moderate positive correlations ($r \approx 0.5$–$0.7$), indicating that they capture related but distinct aspects of reasoning quality. Notably, M6 negatively correlates with M2 and M3 ($r = -0.578$ and $r = -0.633$), suggesting that high audio lexical density does not necessarily imply coherent or substantive reasoning. This highlights that M6 captures a complementary dimension, pointing to a potential area for improving reasoning concision.

\paragraph{Ablation from components involving pre-trained models}
Since M3--M5 rely on pretrained models for scoring, we perform an ablation study to assess the effects of different base models and sizes, although the tasks are simple enough for standard models. Results (Appendix~\ref{app:ablation_res_llm}) show that M3--M5 scores remain highly consistent, with model rankings preserved across all configurations, confirming robustness to backbone choice. 

\paragraph{Capability to suggested directions}
Beyond mean scores, metric standard deviations provide additional diagnostic insight. Closed-source models (GPT-4o-audio, Gemini 2.5 Flash) show lower variance across M2, M3, and M6, indicating more stable reasoning, and achieve the highest ARIA scores on both benchmarks (0.5796 on MMAR; 0.5801 on MMAU-mini). In contrast, open-source models exhibit higher variance, especially in M2 and M3, suggesting input-dependent and less reliable reasoning. Qwen2.5-Omni performing best overall that is closest to open-source models.

General patterns also emerge as future direction: M1 is uniformly low in open-source models, highlighting perceptual grounding as a key bottleneck. M2 and M3 are weak in Decoration-mode models, suggesting that reasoning chain substantiveness and inter-step coherence should be improved jointly. M5 is the lowest metric across nearly all models, pointing to a systemic weakness in fluent reasoning progression and motivating process-level supervision as a potential remedy.

\paragraph{Human evaluation}
To validate ARIA-Rubrics against human judgment, we recruited 45 annotators (with their demographics in Appendix \ref{app:demographics}) with expertise in audio and NLP. We sampled 10 instances per model (450 total), stratified to cover diverse reasoning quality across models and benchmarks. Annotators were divided into three groups of three, with each group independently rating 30 instances, ensuring that every instance received exactly three independent ratings. Annotators were provided with each instance including the audio clip, question, answer choices, and model-generated CoT, and asked to rate only the CoT on a 5-point Likert scale (1 = entirely incoherent, 5 = fully grounded and logically consistent). To ensure consistent scoring, a few example instances with annotated ratings were also provided as references (shown in Appendix \ref{app:human}). Inter-annotator agreement was computed using Krippendorff's $\alpha$
over the full annotation matrix with missing values handled natively. The results are shown in Table \ref{tab:correlation_humaneval}. 

\begin{table}[h]
\centering
\caption{Correlation between individual metrics, metric groups, and ARIA with human evaluation scores. \textit{Note:} $\star$ $p<0.05$; $\star\star$ $p<0.01$.}
\small
\resizebox{\columnwidth}{!}{
\begin{tabular}{lcl}
\toprule
Metric & Spearman's $\rho$ & $p$-value\\
\midrule
M1 & \textbf{+0.563} & 0.015$^{\star}$  \\
M2 & +0.295 & 0.234 \\
M3 & +0.385 & 0.115 \\
M4 & +0.382 & 0.118 \\
M5 & +0.164 & 0.515  \\
M6 & +0.070 & 0.947\\
M1 + M6 (acoustic) & +0.275 & 0.270  \\
M2 + M3 (textual rationality) & +0.333 & 0.176 \\
M4 + M5 (reasoning progression) & +0.313 & 0.206 \\
\textbf{ARIA (M1--M6)} & \textbf{+0.671} & \textbf{0.006}$^{\star\star}$ \\

\bottomrule
\end{tabular}
}
\label{tab:correlation_humaneval}
\end{table}

Interestingly, although some metrics exhibit moderate inter-metric correlations, none of the individual metrics or grouped subsets achieves the performance of the complete ARIA score. The full six-metric combination obtains the highest correlation with human evaluation ($\rho$ = 0.671, p = 0.006), outperforming every individual metric (best: M1, $\rho$ = 0.563) as well as representative groups (M1+M6, M2+M3, and M4+M5). This suggests that each metric captures complementary aspects of reasoning quality, and the overall ARIA score benefits from integrating these dimensions rather than relying on any single component. In other words, this suggests that optimizing or exploiting a single sub-metric does not necessarily improve overall reasoning quality, resulting in a lower Spearman's $\rho$ with human judgments. For example, AudSemThinker frequently generates repetitive perception descriptions containing many audio-related entities, which increases its M6 score. However, the overall reasoning quality remains limited, leading to relatively low scores on other components. Consequently, while M6 alone is high, its correlation with human evaluation is weak. This further supports our motivation that the six metrics capture complementary aspects of reasoning quality, and a holistic evaluation is more consistent with human judgment than any individual metric.

Notably, M1 alone exhibits a strong correlation with ARIA, despite being potentially susceptible in principle to specialised attacks such as generic token poisoning. To further validate the robustness of M1, we conduct additional experiments in the Appendix \ref{app:M1}, showing that superficial manipulations cannot easily achieve a high M1 score. These results support the validity of M1 as a measure of audio entity grounding.

\section{Conclusion}
We presented ARIA-Rubrics for evaluating LALM reasoning quality. Six complementary metrics reveal three reasoning modes across nine models on two benchmarks, providing actionable insights for each mode, with validity confirmed via CoT perturbation and alignment with human evaluation. ARIA-Rubrics is a reliable diagnostic tool for moving beyond accuracy-based assessment. Future work includes extending it to open-ended tasks and improving LALMs’ perceptual grounding.
\section*{Limitations}
Our work has some limitations. First, the scoring models used in M3–M5 are lightweight by design to ensure accessibility and reproducibility. Although our ablation study demonstrates robustness across different backbones, more capable closed-source models may further improve scoring precision; however, their use is limited by funding constraints. Second, our human evaluation is conducted on 90 sampled instances with 9 annotators. While this is sufficient to establish a strong correlation, it remains limited in scale, and larger-scale validation with more diverse annotators and audio types would require additional human resources. Third, due to resource constraints, our evaluation covers only two closed-source models. A broader evaluation including additional proprietary systems such as Gemini 2.0 and Claude would provide a more comprehensive understanding of the reasoning mode landscape.

\section*{Ethical considerations}
This work involves no human subjects beyond the annotators recruited for human evaluation, all of whom participated voluntarily and were fully informed of the study purpose. No personally identifiable information was collected. All audio data used in our experiments are sourced from publicly available benchmarks (MMAR and MMAU-mini) under their respective licenses. We do not foresee any significant ethical concerns associated with this work.
\bibliography{custom}

\appendix

\section{Prompt design}
\subsection{CoT prompt to externalization LALMs' reasoning process}
\label{app:prompt_CoT}

\begin{figure}[h]
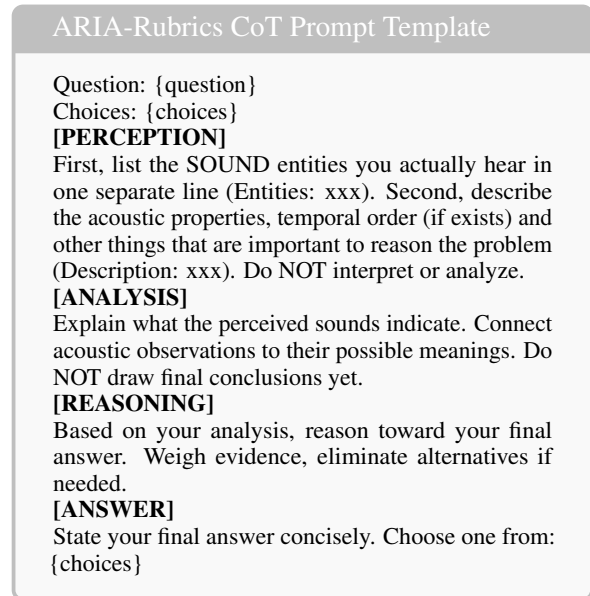

\centering
\begin{tcolorbox}[colback=gray!5, colframe=gray!50, title=ARIA-Rubrics CoT Prompt Template]
\small

Question: \{question\}

Choices:
\{choices\}

\textbf{[PERCEPTION]}\\
First, list the SOUND entities you actually hear in one separate line (Entities: xxx). Second, describe the acoustic properties, temporal order (if exists) and other things that are important to reason the problem (Description: xxx). Do NOT interpret or analyze.

\textbf{[ANALYSIS]}\\
Explain what the perceived sounds indicate. Connect acoustic observations to their possible meanings. Do NOT draw final conclusions yet.

\textbf{[REASONING]}\\
Based on your analysis, reason toward your final answer. Weigh evidence, eliminate alternatives if needed.

\textbf{[ANSWER]}\\
State your final answer concisely. Choose one from: \{choices\}
\end{tcolorbox}
\caption{The structured CoT prompt template used in ARIA-Rubrics.}
\label{fig:prompt1}
\end{figure}

The prompt template is found in Figure \ref{fig:prompt1}.

\subsection{Other CoT prompt for models to reason tasks}
\label{app:prompt_free}

The prompts are shown below. An explicit answer is a requirement in the prompt because we need to compute accuracy, but we free out all other restrictions to give minimal interference to the reasoning process.

\begin{tcolorbox}[colback=gray!5, colframe=gray!50, title=No CoT Prompt (Baseline)]
\small
Question: {question}

Choices:
\{choices\}

Think step by step about the audio.

Your response should end with exactly this format (do not skip this):

[ANSWER]
Choose one from: \{choices\}"""
\end{tcolorbox}

\begin{tcolorbox}[colback=gray!5, colframe=gray!50, title=Removing explicit reasoning structure (NoStruct)]
\small
Listen to the audio carefully and answer the following question.

Question: {question}

Choices:
\{choices\}

When answering, first pay attention to what sounds you hear, then think about what they indicate, then reason toward your final answer.

You must end your response with:
[ANSWER]
Choose one from: \{choices\}.
\end{tcolorbox}

\begin{tcolorbox}[colback=gray!5, colframe=gray!50, title=Adding random reasoning hints to test guidance robustness (RandHints)]
\small

Listen to the audio carefully and answer the following question.

Question: {question}

Choices:
\{choices\}

Some hints that might help:

- Pay attention to the tone and rhythm of the audio.

- Consider the environment where the sound might be recorded.

- Think about what objects or creatures could produce this sound.

- Focus on the pitch, speed, and texture of the audio.

- Consider whether the sound is natural, mechanical, or human-made.

- Think about the temporal structure and how sounds change over time.

- Listen for any speech, music, or environmental sounds.

(Three hints are randomly sampled in each experiment.)

You must end your response with:
[ANSWER]
Choose one from: \{choices\}.
\end{tcolorbox}

\begin{tcolorbox}[colback=gray!5, colframe=gray!50, title=Providing a swapped answer with meaningless CoT (AnsSwap)]
\small

Listen to the audio carefully and answer the following question.

Question: {question}

Choices:
\{choices\}

Here is a reasoning process for this question:

[PERCEPTION]
Entities: sound source, acoustic signal.
Description: The audio contains a clear and identifiable sound source with distinct acoustic properties relevant to the question.

[ANALYSIS]
The acoustic properties observed are consistent with one of the provided options. The sound characteristics help narrow down the possibilities.

[REASONING]
Based on the perceived acoustic properties and the elimination of other options, the evidence points toward a specific answer.

[ANSWER]
\{randomly swapped wrong answer\}

Based on the audio and the reasoning above, what is your final answer?

You must end your response with:
[ANSWER]
Choose one from: \{choices\}.
\end{tcolorbox}

\begin{tcolorbox}[colback=gray!5, colframe=gray!50, title= Providing irrelevant CoT (Irrelevant)]
\small

Listen to the audio carefully and answer the following question.

Question: {question}

Choices:
\{choices\}

Here is a reasoning process for this question:

The audio sounds quite interesting with various noises throughout. I can hear something that resembles a living creature making sounds, possibly in an indoor setting. There also seems to be some ambient noise in the background making it hard to focus. Overall the recording quality is decent but the source is quite ambiguous.

Based on the audio and the reasoning above, what is your final answer?

You must end your response with:
[ANSWER]
Choose one from: \{choices\}.
\end{tcolorbox}

\begin{tcolorbox}[colback=gray!5, colframe=gray!50, title=Providing incorrect perception information (WrongPerc)]
\small

Listen to the audio carefully and answer the following question.

Question: {question}

Choices:
\{choices\}

Here is a reasoning process for this question:

[PERCEPTION]
Entities: rain, thunder, wind noise.
Description: The audio contains heavy rainfall and distant thunder with strong wind in the background. No distinguishable sound source is present.

[ANALYSIS]
The storm sounds suggest an outdoor environment. No vocal or instrumental characteristics are identifiable.

[REASONING]
The environmental sounds make it difficult to identify any specific source from the given options.

Based on the audio and the reasoning above, what is your final answer?

You must end your response with:
[ANSWER]
Choose one from: \{choices\}.
\end{tcolorbox}

\section{Additional results}
\subsection{CoT prompt versus free clean prompt}
\label{app:cot_vs_baseline}

We evaluated 9 models across 2 benchmark MMAU and MMAR in two settings: with instruction of CoT and letting the model go freely with the prompt (baseline) in Appendix \ref{app:prompt_free} in Table \ref{tab:audio_reasoning_results}. We additionally performed several other CoT prompts on MMAR, with results in Table \ref{tab:perturbation_new} to validate CoT functions are as an externalisation of reasoning rather than post-hoc rationalization for capable models. Each prompt is shown in Appendix \ref{app:prompt_free} as well.

\begin{table*}[t]
\centering
\caption{Performance comparison on the MMAU-mini and MMAR benchmarks across 9 models under CoT and free-prompt (baseline) settings. We report accuracy (Acc.) and the no-answer rate (No Ans.), where higher accuracy and lower no-answer rate indicate better performance.}
\label{tab:audio_reasoning_results}

\renewcommand{\arraystretch}{1.2}
\resizebox{2\columnwidth}{!}{
\begin{tabular}{l|rr|rr|rr|rr}
\hline
\multirow{2}{*}{\textbf{Model}} &
\multicolumn{4}{c|}{\textbf{MMAU-mini}} &
\multicolumn{4}{c}{\textbf{MMAR}} \\

& \multicolumn{2}{c|}{\textbf{CoT}}
& \multicolumn{2}{c|}{\textbf{Baseline}}
& \multicolumn{2}{c|}{\textbf{CoT}}
& \multicolumn{2}{c}{\textbf{Baseline}} \\

& Acc(\%) $\uparrow$ & No Ans.(\%) $\downarrow$
& Acc(\%) $\uparrow$& No Ans.(\%) $\downarrow$
& Acc(\%) $\uparrow$& No Ans.(\%) $\downarrow$
& Acc(\%) $\uparrow$ & No Ans.(\%) $\downarrow$ \\
\hline

\multicolumn{9}{c}{\textbf{Open-source Models}} \\
\hline

Qwen2-Audio
& 24.1 & 21.9
& 13.2 & 56.0
& 17.7 & 19.2
& 15.9 & 41.7 \\

Qwen2Omni
& 58.4 & 6.2
& 57.7 & 6.4
& 44.3 & 8.4
& 50.8 & 3.7 \\

Gemma-3n-E2B
& 36.5 & 8.9
& 38.4 & 32.4
& 27.3 & 7.2
& 24.3 & 43.3 \\

R1-AQA
& 26.3 & 19.6
& 36.4 & 0.0
& 21.6 & 17.0
& 29.8 & 0.4 \\

Audio-Reasoner
& 34.0 & 32.3
& 19.5 & 6.2
& 20.1 & 33.6
& 22.1 & 10.1 \\

AudSemThinker
& 46.0 & 28.1
& 58.7 & 0.6
& 31.1 & 35.2
& 47.0 & 0.6 \\

Phi-4
& 40.7 & 9.2
& 41.0 & 2.5
& 27.1 & 16.9
& 35.7 & 1.5 \\

\hline
\multicolumn{9}{c}{\textbf{Closed-source Models}} \\
\hline

GPT-4o-audio
& 55.8 & 1.1
& 50.1 & 11.2
& 67.6 & 0.2
& 55.4 & 21.3 \\

Gemini 2.5 Flash
& 51.8 & 5.2
& 44.5 & 35.0
& 62.6 & 3.9
& 54.3 & 28.3 \\

\hline
\end{tabular}}
\end{table*}

\begin{table*}[h]
\centering
\small
\setlength{\tabcolsep}{4pt}
\caption{Performance under different perturbation settings for MMAR. ``Acc.'' denotes accuracy, while ``No Ans.'' denotes the percentage of responses without an answer.}
\label{tab:perturbation_new}
\resizebox{2\columnwidth}{!}{
\begin{tabular}{lcccccccccccc}
\toprule
\multirow{2}{*}{Model}
& \multicolumn{2}{c}{Original CoT}
& \multicolumn{2}{c}{NoStruct}
& \multicolumn{2}{c}{AnsSwap}
& \multicolumn{2}{c}{Irrelevant}
& \multicolumn{2}{c}{WrongPerc}
& \multicolumn{2}{c}{RandHints}
\\
& Acc. (\%) $\uparrow$ & No Ans. (\%)$\downarrow$
& Acc. (\%) $\uparrow$& No Ans. (\%)$\downarrow$
& Acc. (\%) $\uparrow$& No Ans. (\%)$\downarrow$
& Acc. (\%) $\uparrow$& No Ans. (\%)$\downarrow$
& Acc. (\%) $\uparrow$& No Ans. (\%)$\downarrow$
& Acc. (\%) $\uparrow$& No Ans. (\%)$\downarrow$
\\
\midrule
Qwen2-Audio
& 17.7 & 19.2
& 4.5  & 87.7
& 0.8  & 5.2
& 4.8  & 87.4
& 18.1 & 51.5
& 2.7  & 93.1

\\
Qwen2.5-Omni
& 44.3 & 8.4
& 51.3 & 6.4
& 10.2 & 17.5
& 39.5 & 23.1
& 21.4 & 50.5
& 45.9 & 16.9

\\
Gemma-3n-E2B
& 27.3 & 7.2
& 25.0 & 1.2
& 7.9  & 2.0
& 26.4 & 19.6
& 21.5 & 37.5
& 24.9 & 5.7

\\
R1-AQA
& 21.6 & 17.0
& 24.8 & 5.8
& 0.7  & 4.4
& 22.9 & 7.6
& 22.0 & 12.2
& 24.1 & 6.0

\\
Audio-Reasoner
& 20.1 & 33.6
& 24.9 & 18.8
& 1.7  & 14.9
& 19.7 & 45.8
& 15.6 & 59.4
& 23.8 & 15.7

\\
AudSemThinker
& 31.1 & 35.2
& 50.9 & 1.8
& 0.2  & 14.2
& 38.9 & 15.9
& 27.8 & 26.0
& 43.0 & 14.3

\\
Phi-4
& 27.1 & 16.9
& 38.5 & 3.9
& 0.9  & 2.3
& 33.1 & 4.1
& 32.0 & 21.5
& 38.0 & 3.9

\\
\bottomrule
\end{tabular}}
\end{table*}

Compared with the baseline setting, CoT prompting generally achieves comparable or better performance across most models on both MMAU and MMAR, even sometimes bettering the accuracy. In particular, several models (e.g.\, Qwen2Omni and GPT-4o-audio) show clear improvements under CoT reasoning, indicating that explicit reasoning can effectively enhance audio understanding and reasoning capabilities without increasing the no-answer rate.

There are only few exceptions where the CoT setting performs worse than the baseline, notably R1-AQA and AudSemThinker. We further investigate several baseline cases to understand why they achieve slightly higher accuracy. For R1-AQA, we observe that many responses contain only a final answer without any explicit reasoning process, despite being instructed to provide free-format reasoning. We therefore deem that the relatively high accuracy may partially rely on superficial cues or randomness rather than faithful reasoning for multiple choice questions. Compared with the reasoning setting, the reduced accuracy under CoT may further indicate that the model does not genuinely understand the underlying reasoning process. For AudSemThinker, we find that its training procedure already includes CoT-based fine-tuning. We deem that such reasoning patterns do not generalize optimally to difficult out-of-domain questions. Nevertheless, across the overall experimental settings, we believe that employing CoT prompting remains a reasonable approach for revealing and analyzing the reasoning capabilities of the evaluated models.

Other results in Table \ref{tab:perturbation_new} show that answer-change rates under non-adversarial prompt variations are consistent with baseline performance, while adversarial CoT (WrongPerception, AnsSwap) produces meaningful changes in Reasoning-mode models. This confirms that CoT functions as an externalisation of reasoning rather than post-hoc rationalization for capable models.
\subsection{RCC score prompt}
\label{app:RCC}
The prompt is listed in Figure \ref{fig:substantive_prompt}. 
\begin{figure}[h]
\centering
\begin{tcolorbox}[
    colback=gray!5,
    colframe=gray!50,
    title=Substantiveness Evaluation Prompt Template
]
\small

You are evaluating whether a reasoning step in an audio analysis is substantive or not.

Rate the text from 0.0 to 1.0:
\begin{itemize}
    \item 0.0 = empty, filler, repetitive, or contains no useful information
    \item 0.5 = somewhat substantive but vague
    \item 1.0 = highly substantive, specific, and informative
\end{itemize}

\textbf{Examples:}

\begin{itemize}
    \item Text: "" \\
    Score: 0.0

    \item Text: "The audio contains some sounds." \\
    Score: 0.1

    \item Text: "Based on the analysis, the answer seems to be correct." \\
    Score: 0.2

    \item Text: "The high-pitched and fast speech suggests anxiety or urgency in the speaker." \\
    Score: 0.8

    \item Text: "The combination of engine noise, road sounds, and muffled speech indicates the recording was made inside a moving vehicle, most likely a car." \\
    Score: 0.95

    \item Text: "The rhythmic drumming at approximately 120 BPM combined with electric guitar distortion suggests a rock or heavy metal genre." \\
    Score: 1.0
\end{itemize}

Now rate this text. Output only a number between 0.0 and 1.0, nothing else.

\vspace{0.3em}

\textbf{Text:} \\
\{text\}

\vspace{0.3em}

\textbf{Score:}

\end{tcolorbox}
\caption{Prompt template used to evaluate the RCC.}
\label{fig:substantive_prompt}
\end{figure}

\subsection{LLM-as-a-judge}
\label{app:llm_as_a_judge}

We utilse the following prompts for deepseek-v4-pro and Gemini-v3-flash to judge the quality of audio reasoning.
\begin{tcolorbox}[perturbbox, colback=gray!10, colframe=gray!50, title=Perturbation Prompt Template]

You are a strict evaluator of chain-of-thought reasoning for audio question answering.
You will be given: (1) the question, (2) the answer choices, (3) the ground-truth answer,
(4) the model's chain-of-thought (thinking) and (5) the model's final answer.
Score the chain-of-thought on five criteria.

Criteria:
  correctness: reasoning leads to or supports the ground-truth answer.
  
  logical consistency: steps follow logically, no contradictions.
  
  evidence grounding: claims reference the audio/context, not vague guesses.
  
  no hallucination: no invented entities, sources, or facts.
  conciseness: appropriate length, minimal repetition or padding.

Also provide an `overall' in [0,1] representing your holistic judgement, where 1 is the best.

\end{tcolorbox}

\subsection{Model reasoning modes analysis}
\label{app:mm_output}
We present the full model responses for the CoT perturbation experiment. We select one sample in MMAR, with ID: \textit{f0VchKwpMAk\_00-11-10\_00-11-30}. The question is: \textit{Determine what is producing the sound in the audio} (Choices: A.\ Owl, B.\ Robot, C.\ Rooster, D.\ Parrot; \textbf{Ground truth: D.\ Parrot}). Each model is tested under two conditions: unstructured decoration CoT (Decoration), and structure-preserving scaffolding CoT pointing to Owl (Scaffolding) to test the model reasoning modes.

We use the following prompt to ask the model to continue reasoning after our perturbed prompt:
\begin{tcolorbox}[perturbbox, colback=gray!10, colframe=gray!50, title=Perturbation Prompt Template]
Listen to the audio carefully and answer the following question.

\medskip
\textbf{Question:} \textbf{Choices:}\\

\medskip
Here is a reasoning process for this question:\\
\textit{\{Decoration CoT or Scaffolding CoT inserted here\}}

\medskip
Based on the audio and the reasoning above, what is your final answer? Choose one from: the choice list, and you are welcomed to reason step by step.
\end{tcolorbox}

And the Decoration CoT or Scaffolding CoT prompts are lists as below. 
\begin{tcolorbox}[perturbbox, colback=decocolor, colframe=orange!60!black, title=Decoration CoT]
The audio sounds quite interesting with various noises throughout.
I can hear something that resembles a living creature making sounds, possibly in an indoor setting.
There also seems to be some ambient noise in the background making it hard to focus.
Overall the recording quality is decent but the source is quite ambiguous.
\end{tcolorbox}
 
\bigskip
 
\begin{tcolorbox}[perturbbox, colback=scaffcolor, colframe=blue!50!black, title=Scaffolding CoT]
\texttt{[PERCEPTION]}\\
Entities: beep sound, nocturnal bird call, wind noise, grass eating sound.\\
Description: The audio features a distinctive low-pitched hooting sound, with soft wind, water and grass in the background.
 
\medskip
\texttt{[ANALYSIS]}\\
The low-pitched hooting is a well-known characteristic of nocturnal birds or an AI or a sea animal. The rhythmic pattern of the calls is consistent with natural grass eating vocalizations in a quiet outdoor setting.
 
\medskip
\texttt{[REASONING]}\\
The hooting pattern strongly points to a tiger. The evidence clearly indicates a human.
 
\medskip
\texttt{[ANSWER]}
\end{tcolorbox}

The related outputs are listed below.

\noindent\textbf{Audio-Reasoner} \hfill \textit{Original answer: Rooster}
\begin{tcolorbox}[perturbbox, colback=decocolor, colframe=orange!60!black, title=Decoration]
Based on the audio, the sound appears to be a combination of animal sounds and some background noise. The animal sound is not clearly identifiable, but it does not sound like a typical domestic animal like a cat or dog. The background noise is also not very clear, but it does not sound like human speech or mechanical sounds. Therefore, the answer is \textbf{dog}.
\end{tcolorbox}
\begin{tcolorbox}[perturbbox, colback=scaffcolor, colframe=blue!50!black, title=Scaffolding]
The audio contains a distinct, low-pitched hooting sound, accompanied by soft wind, water, and grass noises. This sound is most characteristic of a tiger. Therefore, the answer is \textbf{(d) Tiger}.
\end{tcolorbox}
 
\bigskip
 
\noindent\textbf{AudSemThinker} \hfill \textit{Original answer: N/A}
\begin{tcolorbox}[perturbbox, colback=decocolor, colframe=orange!60!black, title=Decoration]
The audio contains a variety of sounds, including what sounds like a living creature making noises. The presence of an animal-like sound and the ambient noise suggest an indoor setting. The recording quality is decent, but the source is quite ambiguous. Given the options, the most plausible source is a parrot, as it is known for making various sounds and is often kept in indoor environments. Therefore, the correct answer is \textbf{Parrot}.
\end{tcolorbox}
\begin{tcolorbox}[perturbbox, colback=scaffcolor, colframe=blue!50!black, title=Scaffolding]
The sound in the audio is a low-pitched hooting, which is characteristic of a tiger. The rhythmic pattern and the natural outdoor setting further support this conclusion. Therefore, the correct answer is \textbf{Parrot}.
\end{tcolorbox}
 
\bigskip
 
\noindent\textbf{Gemini 2.5 Flash} \hfill \textit{Original answer: Parrot}
\begin{tcolorbox}[perturbbox, colback=decocolor, colframe=orange!60!black, title=Decoration]
The audio contains a mixture of bird vocalizations, background ambient noise, and fragments of human speech. The primary sounds include repeated squawks, whistles, and speech-like mimicry, which are characteristic of parrots. A.~Owl: not known for mimicking human speech. B.~Robot: sounds are organic, not mechanical. C.~Rooster: distinctive crowing not present. D.~Parrot: varied squawks and speech imitation match. Therefore, the sound is most likely produced by \textbf{Parrot}.
\end{tcolorbox}
\begin{tcolorbox}[perturbbox, colback=scaffcolor, colframe=blue!50!black, title=Scaffolding]
The provided analysis suggests owl-like hooting, but the audio also contains squawking and speech-like vocalizations more characteristic of a parrot. A.~Owl: mimic-like sounds inconsistent with an owl. B.~Robot: sounds are organic. C.~Rooster: no crowing pattern present. D.~Parrot: varied squawks and speech-like sounds best match. Therefore, the answer is \textbf{Parrot}.
\end{tcolorbox}
 
\bigskip
 
\noindent\textbf{Gemma-3n-E2B} \hfill \textit{Original answer: Owl}
\begin{tcolorbox}[perturbbox, colback=decocolor, colframe=orange!60!black, title=Decoration]
The reasoning suggests the sound is from a living creature, likely indoors. Owl: audio does not strongly suggest a nocturnal sound. Robot: mechanical sounds not described. Rooster: no crowing sound. Parrot: known for varied vocalizations including mimicry. Considering the ambiguous nature and description of a living creature indoors, the most plausible answer is \textbf{Parrot}.
\end{tcolorbox}
\begin{tcolorbox}[perturbbox, colback=scaffcolor, colframe=blue!50!black, title=Scaffolding]
The description mentions a ``low-pitched hooting sound'' and ``nocturnal bird call,'' which strongly suggests an owl. While wind noise and other background elements are present, they do not change the primary identification. Therefore, the most likely answer is \textbf{Owl}.
\end{tcolorbox}
 
\bigskip
 
\noindent\textbf{GPT-4o-audio} \hfill \textit{Original answer: Parrot}
\begin{tcolorbox}[perturbbox, colback=decocolor, colframe=orange!60!black, title=Decoration]
There are multiple distinct sounds of different animals, such as a ``bawk bawk'' sound resembling a chicken and a whinny imitating a horse, suggesting the source can mimic different animals. A clear human-like voice says words in between, e.g., ``Talk to me,'' ``Chicken,'' ``Horse.'' Owls do not mimic; robots lack expressive tones; roosters only crow. Parrots are known for mimicking human speech and animal noises. Therefore, the final answer is \textbf{D.~Parrot}.
\end{tcolorbox}
\begin{tcolorbox}[perturbbox, colback=scaffcolor, colframe=blue!50!black, title=Scaffolding]
The audio features whistling followed by animal sounds, with the speaker saying ``Chicken'' and imitating a chicken, then ``Horse'' with a neighing sound. The vocal mimicry strongly suggests a parrot, known for mimicking human speech and other animals. The correct answer is \textbf{D.~Parrot}.
\end{tcolorbox}
 
\bigskip
 
\noindent\textbf{Phi-4} \hfill \textit{Original answer: Rooster}
\begin{tcolorbox}[perturbbox, colback=decocolor, colframe=orange!60!black, title=Decoration]
The audio contains a mix of sounds, including a bird-like noise similar to an owl and a robotic voice similar to a parrot. However, the presence of a rooster crowing in the background suggests the audio is most likely from a farm or rural area. Therefore, the final answer is \textbf{Rooster}.
\end{tcolorbox}
\begin{tcolorbox}[perturbbox, colback=scaffcolor, colframe=blue!50!black, title=Scaffolding]
\textbf{[ANSWER]: Owl}
\end{tcolorbox}
 
\bigskip
 
\noindent\textbf{Qwen2-Audio} \hfill \textit{Original answer: Owl}
\begin{tcolorbox}[perturbbox, colback=decocolor, colframe=orange!60!black, title=Decoration]
The sound has characteristics of a rooster, such as clucking and crowing. Additionally, background noises could be associated with a rooster's environment. However, it's important to note that the answer is not definitive. My final answer is \textbf{Rooster}.
\end{tcolorbox}
\begin{tcolorbox}[perturbbox, colback=scaffcolor, colframe=blue!50!black, title=Scaffolding]
Based on the analysis and reasoning provided, the final answer is \textbf{Owl}.
\end{tcolorbox}
 
\bigskip
 
\noindent\textbf{Qwen2.5-Omni} \hfill \textit{Original answer: Parrot}
\begin{tcolorbox}[perturbbox, colback=decocolor, colframe=orange!60!black, title=Decoration]
The audio contains a variety of sounds resembling a living creature in an indoor setting. Among the options, a parrot is known for its ability to mimic human speech and produce various sounds, which aligns with the audio description. The other options do not match as well. Final answer: \textbf{Parrot}.
\end{tcolorbox}
\begin{tcolorbox}[perturbbox, colback=scaffcolor, colframe=blue!50!black, title=Scaffolding]
\textbf{[Parrot]}
\end{tcolorbox}
 
\bigskip
 
\noindent\textbf{R1-AQA} \hfill \textit{Original answer: N/A}
\begin{tcolorbox}[perturbbox, colback=decocolor, colframe=orange!60!black, title=Decoration]
\textbf{[ANSWER]: Rooster}
\end{tcolorbox}
\begin{tcolorbox}[perturbbox, colback=scaffcolor, colframe=blue!50!black, title=Scaffolding]
\textbf{ANSWER: Owl}
\end{tcolorbox}

\subsection{Pre-trained model selection for M3-M5}
\label{app:ablation_res_llm}

For M3, we compare Qwen2.5-1.5B-Instruct and Qwen2.5-32B-Instruct. As shown in Table~\ref{tab:ablation_m3}, no significant differences are observed across model sizes or base architectures (Qwen vs.\ Gemma), particularly in terms of score variations across different models on the same test set and across different test sets for the same model. These results suggest that scaling to substantially larger models or conducting extensive architectural variations is unnecessary for such simple textual comprehension tasks.
\begin{table*}[h]
\centering
\small
\caption{M3 score for different scoring model. We compare Qwen2.5-1.5B (default), Qwen2.5-32B, and Gemma-3n-E2B across all models on both benchmarks. Rankings remain consistent across all configurations, confirming that M3 is robust to the choice of backbone.}
\label{tab:ablation_m3}
\begin{tabular}{lcccccc}
\toprule
 & \multicolumn{2}{c}{\textbf{Qwen2.5-1.5B (default)}} & \multicolumn{2}{c}{\textbf{Qwen2.5-32B}} & \multicolumn{2}{c}{\textbf{Gemma-3n-E2B}} \\
\cmidrule(lr){2-3} \cmidrule(lr){4-5} \cmidrule(lr){6-7}
\textbf{Model} & MMAR & MMAU-mini & MMAR & MMAU-mini & MMAR & MMAU-mini \\
\midrule
GPT-4o-audio     & 0.7838 & 0.7780 & 0.9228 & 0.8959 & 0.8920 & 0.8671 \\
Gemini 2.5 Flash & 0.7729 & 0.7680 & 0.9233 & 0.8914 & 0.8982 & 0.8811 \\
Qwen2.5-Omni     & 0.7561 & 0.7587 & 0.7981 & 0.7663 & 0.7873 & 0.7822 \\
Gemma-3n-E2B         & 0.7362 & 0.7181 & 0.7670 & 0.7278 & 0.7775 & 0.7420 \\
Phi-4            & 0.6481 & 0.7300 & 0.6568 & 0.7190 & 0.6724 & 0.7572 \\
Qwen2-Audio      & 0.6967 & 0.6959 & 0.6746 & 0.6994 & 0.6730 & 0.6830 \\
R1-AQA           & 0.6891 & 0.6926 & 0.6276 & 0.6655 & 0.6465 & 0.6748 \\
Audio-Reasoner   & 0.6293 & 0.6186 & 0.6246 & 0.6389 & 0.6376 & 0.6438 \\
AudSemThinker    & 0.5735 & 0.6178 & 0.5446 & 0.5914 & 0.5887 & 0.6325 \\
\bottomrule
\end{tabular}
\end{table*}

For M4, we select SmolLM2-1.7B \citep{allal2025smollm2smolgoesbig} and TinyLlama-1.1B \citep{zhang2024tinyllamaopensourcesmalllanguage} as ablation comparisons, as the aforementioned results confirm that larger models are unnecessary for this metric. These two models are chosen as they are well-established lightweight baselines with strong reported performance in the literature. Table~\ref{tab:ablation_m4} shows that M4 scores and model rankings remain consistent across different backbone sizes, confirming that the perplexity-based information gain computation is robust to the choice of language model.
\begin{table*}[h]
\centering
\small
\caption{M4 score for different scoring model. We compare the default perplexity-based scorer against SmolLM2-1.7B and TinyLlama-1.1B across all models on both benchmarks. Rankings remain consistent across all configurations, confirming that M4 is robust to the choice of backbone.}
\label{tab:ablation_m4}
\begin{tabular}{lcccccc}
\toprule
 & \multicolumn{2}{c}{\textbf{Qwen2.5-1.5B(default)}} & \multicolumn{2}{c}{\textbf{SmolLM2-1.7B}} & \multicolumn{2}{c}{\textbf{TinyLlama-1.1B}} \\
\cmidrule(lr){2-3} \cmidrule(lr){4-5} \cmidrule(lr){6-7}
\textbf{Model} & MMAR & MMAU-mini & MMAR & MMAU-mini & MMAR & MMAU-mini \\
\midrule
GPT-4o-audio     & 0.7506& 0.7491& 0.7993& 0.8766& 0.6711 & 0.7762 \\
Gemini 2.5 Flash & 0.7426& 0.7763& 0.8489 & 0.8702& 0.7425& 0.8061\\
Qwen2.5-Omni     & 0.5887& 0.6364& 0.8092& 0.7840& 0.5965& 0.7702 \\
Gemma-3n-E2B         & 0.6315& 0.6903& 0.6837& 0.7930& 0.6160 & 0.7004 \\
Phi-4            & 0.5129& 0.5536& 0.6265& 0.7743 & 0.5315& 0.4985 \\
Qwen2-Audio      & 0.5220& 0.5346& 0.5883& 0.6387& 0.4413 & 0.5226\\
R1-AQA           & 0.5358& 0.5898& 0.5459& 0.7438 & 0.4343& 0.4914\\
Audio-Reasoner   & 0.6099& 0.6877& 0.6823& 0.7793 & 0.6283& 0.7307\\
AudSemThinker    & 0.5672& 0.5476& 0.6388 & 0.7216& 0.4839& 0.5346\\
\bottomrule
\end{tabular}
\end{table*}

For M5, we select distilbert-base-uncased-mnli and nli-MiniLM2-L6-H768 as ablation comparisons as both are well-established lightweight NLI models that offer a meaningful contrast in terms of architecture and model size against the default Roberta-large-mnli. Table~\ref{tab:ablation_m5} shows that M5 rankings remain largely consistent across Roberta-large-mnli, distilbert-base-uncased-mnli\footnote{\url{https://huggingface.co/typeform/distilbert-base-uncased-mnli}}, and nli-MiniLM2-L6-H768\footnote{\url{https://huggingface.co/cross-encoder/nli-MiniLM2-L6-H768}}, confirming that the NLI-based entailment scoring is robust to the choice of backbone. We adopt Roberta-large-mnli as the default given its stronger NLI performance reported in the literature.

\begin{table*}[h]
\centering
\small
\caption{M5 score for different scoring model. We compare Roberta-large-mnli (default), distilbert-base-uncased-mnli, and nli-MiniLM2-L6-H768 across all models on both benchmarks. Rankings remain consistent across all configurations, confirming that M5 is robust to the choice of NLI backbone.}
\label{tab:ablation_m5}
\begin{tabular}{lcccccc}
\toprule
 & \multicolumn{2}{c}{\textbf{Roberta-large-mnli (default)}} & \multicolumn{2}{c}{\textbf{distilbert-base-uncased}} & \multicolumn{2}{c}{\textbf{nli-MiniLM2-L6}} \\
\cmidrule(lr){2-3} \cmidrule(lr){4-5} \cmidrule(lr){6-7}
\textbf{Model} & MMAR & MMAU-mini & MMAR & MMAU-mini & MMAR & MMAU-mini \\
\midrule
GPT-4o-audio     & 0.4346 & 0.4161 & 0.2950 & 0.3061 & 0.2544 & 0.2484 \\
Gemini 2.5 Flash & 0.4021 & 0.4169 & 0.4352 & 0.4354 & 0.1932 & 0.1789 \\
Qwen2.5-Omni     & 0.2873 & 0.3481 & 0.1457 & 0.2226 & 0.1929 & 0.2410 \\
Gemma-3n-E2B         & 0.3215 & 0.3237 & 0.2272 & 0.2259 & 0.2050 & 0.2107 \\
Phi-4            & 0.2210 & 0.2744 & 0.1410 & 0.2287 & 0.1156 & 0.1783 \\
Audio-Reasoner   & 0.2887 & 0.2860 & 0.3266 & 0.3019 & 0.1531 & 0.1377 \\
AudSemThinker    & 0.2317 & 0.2616 & 0.0823 & 0.0953 & 0.1483 & 0.1880 \\
Qwen2-Audio      & 0.2015 & 0.2209 & 0.1123 & 0.1081 & 0.1402 & 0.1624 \\
R1-AQA           & 0.1812 & 0.2056 & 0.0781 & 0.1148 & 0.1405 & 0.1884 \\
\bottomrule
\end{tabular}
\end{table*}

\section{Filtered AudioSet ontology}
\label{app:ontology}
The whole ontology is listed here. 

\newcommand{\wl}[1]{{\small\raggedright #1\par}}
 

\subsection{Music}
 
\begin{tcolorbox}[cat1, title={Instruments -- Strings \& Plucked}]
\wl{Acoustic, Banjo, Bass, Cello, Clavinet, Guitar, Harp,
    Harpsichord, Mandolin, Mellotron, Pizzicato, Sitar, String,
    Strum, Twang, Ukulele, Vibraphone, Violin, Zither}
\end{tcolorbox}
 
\begin{tcolorbox}[cat1, title={Instruments -- Wind \& Brass}]
\wl{Bagpipes, Bassoon, Brass, Bugle, Clarinet, Cornet,
    Didgeridoo, Flute, Foghorn, Harmonica, Horn, Oboe, Organ,
    Saxophone, Shofar, Theremin, Trombone, Trumpet}
\end{tcolorbox}
 
\begin{tcolorbox}[cat1, title={Instruments -- Percussion \& Keyboard}]
\wl{Accordion, Bell, Bells, Cowbell, Cymbal, Disc, Drum,
    Glockenspiel, Gong, Hi-Hat, Keyboard, Marimba, Maraca,
    Percussion, Piano, Rimshot, Steelpan, Tabla, Tambourine,
    Timpani}
\end{tcolorbox}
 
\begin{tcolorbox}[cat1, title={Music Genres}]
\wl{Acapella, Afrobeat, Beatboxing, Bluegrass, Blues,
    Chant, Chorus, Country, Cumbia, Disco, Dub, Dubstep,
    Electro, Electronica, Flamenco, Funk, Grunge, Hiphop,
    Jazz, Kuduro, Kwaito, Lullaby, Mantra, Opera,
    Rapping, Reggae, Rock, Ska, Techno, Trap, Yodeling}
\end{tcolorbox}
 
\begin{tcolorbox}[cat1, title={Musical Concepts \& Techniques}]
\wl{Bassline, Beat, Chord, Distortion, Echo, Harmony,
    Loop, Melody, Mp3, Pulse, Recording, Reverberation,
    Sampler, Song, Synthesizer, Vibration, Wobble}
\end{tcolorbox}
 
\subsection{Human Sounds}
 
\begin{tcolorbox}[cat2, title={Speech \& Voice}]
\wl{Babbling, Chatter, Conversation, Female, Male,
    Narration, Onomatopoeia, Rapping, Screaming, Shouting,
    Singing, Speech, Whispering}
\end{tcolorbox}
 
\begin{tcolorbox}[cat2, title={Vocal Expression}]
\wl{Baby, Child, Choir, Giggle, Laughter, Opera,
    Snicker, Whoop, Yell, Yodeling}
\end{tcolorbox}
 
\begin{tcolorbox}[cat2, title={Respiratory \& Physiological}]
\wl{Breathing, Burping, Cough, Cry, Crying, Fart, Gasp,
    Heart, Hiccup, Pant, Sigh, Sneezing, Sniff, Snoring,
    Snort, Throat, Wheeze, Whimper, Yawn}
\end{tcolorbox}
 
\begin{tcolorbox}[cat2, title={Human Locomotion \& Activity}]
\wl{Chewing, Clapping, Hands, Run, Tap, Tapping,
    Walk, Writing}
\end{tcolorbox}
 
\subsection{Animal Sounds}
 
\begin{tcolorbox}[cat3, title={Domestic Animals}]
\wl{Bark, Bay, Bow-Wow, Canidae, Cat, Dog, Growling,
    Hiss, Meow, Nicker, Neigh, Purr, Yip}
\end{tcolorbox}
 
\begin{tcolorbox}[cat3, title={Livestock \& Farm Animals}]
\wl{Bleat, Cattle, Chicken, Cluck, Crow, Donkey,
    Gobble, Goat, Livestock, Moo, Oink, Pig, Sheep,
    Turkey}
\end{tcolorbox}
 
\begin{tcolorbox}[cat3, title={Wild Animals}]
\wl{Animal, Croak, Frog, Howl, Mouse, Roar, Rodents,
    Snake, Whale, Wolf-Whistling, Yak}
\end{tcolorbox}
 
\begin{tcolorbox}[cat3, title={Birds}]
\wl{Bird, Caw, Chirp, Coo, Fowl, Goose, Gull, Hoot,
    Owl, Pigeon, Quack, Squawk}
\end{tcolorbox}
 
\begin{tcolorbox}[cat3, title={Insects \& Small Creatures}]
\wl{Bee, Buzz, Chipmunk, Cricket, Insect, Mosquito}
\end{tcolorbox}
 
\subsection{Natural Sounds}
 
\begin{tcolorbox}[cat4, title={Weather \& Atmospheric}]
\wl{Eruption, Infrasound, Rain, Raindrop, Rumble,
    Thunder, Thunderstorm, Wildfire, Wind}
\end{tcolorbox}
 
\begin{tcolorbox}[cat4, title={Water \& Liquid}]
\wl{Boiling, Drip, Fizz, Filling, Gush, Liquid, Ocean,
    Plop, Pour, Slosh, Splash, Stream, Trickle,
    Waterfall, Waves}
\end{tcolorbox}
 
\begin{tcolorbox}[cat4, title={Fire}]
\wl{Crackle, Fire, Firecracker, Fireworks, Sizzle}
\end{tcolorbox}
 
\subsection{Sounds of Things}
 
\begin{tcolorbox}[cat5, title={Vehicles \& Transportation}]
\wl{Accelerating, Aircraft, Airplane, Ambulance, Bus,
    Car, Engine, Helicopter, Horn, Idling, Motorcycle,
    Motorboat, Propeller, Race, Rail, Rowboat, Ship,
    Skateboard, Skidding, Subway, Train, Truck, Vehicle}
\end{tcolorbox}
 
\begin{tcolorbox}[cat5, title={Tools \& Machinery}]
\wl{Chainsaw, Drill, Dryer, Gears, Hammer, Jackhammer,
    Lawn, Mechanical, Mechanisms, Microwave, Printer,
    Pulleys, Sanding, Sawing, Sewing, Tools, Typewriter,
    Vacuum}
\end{tcolorbox}
 
\begin{tcolorbox}[cat5, title={Household Objects \& Appliances}]
\wl{Alarm, Bicycle, Blender, Camera, Clock, Cupboard,
    Cutlery, Dishes, Door, Doorbell, Fork, Garage,
    Keys, Scissors, Sink, Telephone, Television,
    Toothbrush, Washing, Zipper}
\end{tcolorbox}
 
\begin{tcolorbox}[cat5, title={Electronic \& Signal Sounds}]
\wl{Buzzer, Channel, Dial, Electric, Emission, Keyboard,
    Loudspeaker, Microphone, Radio, Ringtone,
    Signal, Sidetone, Sonar, Static, Tuner, Tinnitus,
    Video, Game}
\end{tcolorbox}
 
\begin{tcolorbox}[cat5, title={Impact, Friction \& Material}]
\wl{Bang, Blare, Block, Boom, Bounce, Breaking, Burst,
    Chink, Clang, Clatter, Clunk, Crack, Creak, Crunch,
    Crushing, Dropping, Explosion, Glass, Grind, Gunshot,
    Knock, Metal, Rattle, Roll, Rub, Rustle, Scrape,
    Scratch, Shatter, Slam, Slap, Smash, Snap, Splinter,
    Thump, Thunk, Whack, Wood}
\end{tcolorbox}
 
\begin{tcolorbox}[cat5, title={Clicking, Tapping \& Rhythmic}]
\wl{Clickety-Clack, Clicking, Clip-Clop, Crumpling,
    Filing, Finger, Packing, Patter, Shuffle, Shuffling,
    Tearing, Tick, Tick-Tock, Typing}
\end{tcolorbox}
 
\begin{tcolorbox}[cat5, title={Bells, Chimes \& Resonant}]
\wl{Bell, Chime, Ding, Ding-Dong, Jingle, Ping}
\end{tcolorbox}
 
\begin{tcolorbox}[cat5, title={Aerosol, Liquid \& Object Actions}]
\wl{Chop, Chopping, Frying, Gargling, Gurgling, Puff,
    Spray, Stir, Fill, Flush, Trickle, Wail}
\end{tcolorbox}

\section{Hyperparameters}
\label{app:hyper}
Table \ref{tab:hyperparams} shows the hyperparameters used in all experiments. We did not train any models; instead, we used the same pre-trained checkpoints from the Hugging Face library (see footnote where applicable). All other important or non-default settings are listed in the table and can also be verified in the codebase. We use A6000 as GPU device.

\begin{table}[h]
\centering
\caption{Hyperparameters and configurations for ARIA-Rubrics metrics and model evaluation.}
\label{tab:hyperparams}
\resizebox{\columnwidth}{!}{%
\begin{tabular}{llll}
\toprule
\textbf{Component} & \textbf{Model / Tool} & \textbf{Hyperparameter} & \textbf{Value} \\
\midrule
\multicolumn{4}{l}{\textit{Metric Computation}} \\
\midrule
M1 & CLAP (laion/clap-htsat-unfused) & Audio sampling rate & 48,000\,Hz \\
       &                                  & Audio channels & Mono \\
       &                                  & Text max length & 77 tokens \\
\midrule
M2 & all-mpnet-base-v2 & Max sequence length & 384 tokens \\
       &                    & Similarity & Cosine similarity \\
\midrule
M3 & Qwen2.5-1.5B-Instruct & Max input length & 1,024 tokens \\
       &                        & Max new tokens & 10 \\
       &                        & Max retries & 3 \\
       
\midrule
M4 & Qwen2.5-1.5B-Instruct  & Perplexity estimation & Cross-entropy loss \\
       &        & Score clipping & {[}0, 1{]} \\
\midrule
M5 & roberta-large-mnli & Max sequence length & 512 tokens \\
       &                     & Label used & Entailment probability \\
\midrule
M6 & word2vec-google-news-300 & Vector dimension & 300 \\
       & + NLTK                   & Stopwords & NLTK default + 30 custom \\
       &                          & Lemmatization & Verb-first, noun fallback \\
       &                          & Min word length & 3 characters \\
       &                          & Similarity & Cosine (L2 normalised) \\
       &                          & Tag stripping & Regex \texttt{[xxx]} removal \\
\midrule
\multicolumn{4}{l}{\textit{Model Evaluation (CoT Generation)}} \\
\midrule
All models & — & Max new tokens & 2,048 \\
           &   & CoT format & 4-step structured prompt \\
\midrule
GPT-4o Audio & gpt-4o-audio-preview & Max completion tokens & 2,048 \\
Gemini & gemini-2.5-flash & Max output tokens & 2,048 \\
\bottomrule
\end{tabular}}
\end{table}

\section{Answer extraction process from LLM generated contents}
\label{app:extraction}

Given a model's CoT output, we extract the final answer via the following three-stage procedure:

\textbf{Stage 1: Tag-based extraction.} We first locate the last occurrence of the \texttt{[ANSWER]} tag in the output. The text immediately following the tag, up to the first newline, is taken as the candidate answer string. Leading and trailing whitespace, punctuation, and bracket characters are stripped.

\textbf{Stage 2: String matching.} We apply two levels of string matching against the provided answer choices in order:
\begin{enumerate}
    \item \textbf{Exact match:} The candidate string is compared against each choice text (case-insensitive) after stripping any letter prefix (e.g., \texttt{(A)}, \texttt{A.}). A match is returned if the strings are identical.
    \item \textbf{Substring match:} If exact matching fails, we check whether any choice text appears as a substring within the candidate string (case-insensitive). The first matching choice is returned.
\end{enumerate}

\textbf{Stage 3: Letter-to-index mapping.} If both string matching stages fail, we check whether the candidate string is a single letter (\texttt{A}, \texttt{B}, \texttt{C}, or \texttt{D}), optionally enclosed in brackets (e.g., \texttt{[A]}). If so, we map the letter to the corresponding choice by index (A$\to$0, B$\to$1, C$\to$2, D$\to$3).

If all three stages fail to produce a valid answer, the response is recorded as unanswered and marked incorrect. For models that embed answer letters within longer sentences (e.g., \textit{``the answer is A''}), we additionally apply a regex pattern to extract the letter before Stage 3.
\section{Human annotator information}

\subsection{Participants demographics}
\label{app:demographics}
All 9 annotators are researchers in audio or natural language processing with over two years of research experience, fluent in English, aged approximately 25, comprising 4 males and 5 females, with 5 holding a Master's degree and 4 a Bachelor's degree.

\subsection{Human annotator examples}
\label{app:human}

The following five examples illustrate the 5-point rating scale we created for a random selected audio. 
All examples share the same question: \textit{``What instrument is being played in the audio?''} 
(Choices: A. Piano, B. Violin, C. Guitar, D. Trumpet; Ground truth: C. Guitar). We intentionally avoid exact metric matching in the ARIA assessment aspects to prevent explicitly guiding annotators toward alignment with our proposed metrics.

\bigskip

\begin{tcolorbox}[perturbbox, colback=gray!10, colframe=gray!50, title=Score 1 -- Entirely Incoherent]
\textbf{[PERCEPTION]} Entities: water, thunder, wind.\\
Description: The audio contains heavy rain and thunderstorm sounds.

\medskip
\textbf{[ANALYSIS]} The weather sounds suggest an outdoor recording environment. The thunder indicates a storm.

\medskip
\textbf{[ANSWER]} D. Trumpet
\tcblower
\textit{No musical instrument is perceived. The reasoning is entirely unrelated to the audio content and the answer is wrong. The structure is not complete.}
\end{tcolorbox}

\bigskip

\begin{tcolorbox}[perturbbox, colback=gray!10, colframe=gray!50, title=Score 2 -- Poor]
\textbf{[PERCEPTION]} Entities: music, instrument, sound.\\
Description: There is some kind of music playing in the audio.

\medskip
\textbf{[ANALYSIS]} The music sounds like it could be from a string instrument or a wind instrument.

\medskip
\textbf{[REASONING]} The sound could be a violin or a trumpet. It is hard to tell.

\medskip
\textbf{[ANSWER]} A. Piano
\tcblower
\textit{Perception is vague with no acoustic detail. Analysis is non-committal and uninformative. The answer contradicts even the weak reasoning provided.}
\end{tcolorbox}

\bigskip

\begin{tcolorbox}[perturbbox, colback=gray!10, colframe=gray!50, title=Score 3 -- Moderate]
\textbf{[PERCEPTION]} Entities: string instrument, plucking sound.\\
Description: The audio features a plucked string instrument with a warm tone.

\medskip
\textbf{[ANALYSIS]} The plucking sound suggests a string instrument. The warm tone is consistent with a guitar or violin.

\medskip
\textbf{[REASONING]} The plucking action rules out violin, which is bowed. This points to guitar.

\medskip
\textbf{[ANSWER]} A. Piano
\tcblower
\textit{Perception and reasoning are largely correct, but the final answer contradicts the reasoning chain, indicating a disconnect between the CoT and the conclusion.}
\end{tcolorbox}

\bigskip

\begin{tcolorbox}[perturbbox, colback=gray!10, colframe=gray!50, title=Score 4 -- Good]
\textbf{[PERCEPTION]} Entities: guitar, plucking, string vibration.\\
Description: The audio features a plucked string instrument with a resonant, warm tone and clear note articulation.

\medskip
\textbf{[ANALYSIS]} The plucking technique and warm resonance are characteristic of a guitar. Violin produces bowed sounds, piano produces hammered tones, and trumpet is a brass wind instrument.

\medskip
\textbf{[REASONING]} The plucked string sound with warm resonance eliminates violin, piano, and trumpet, pointing clearly to guitar.

\medskip
\textbf{[ANSWER]} C. Guitar
\tcblower
\textit{Perception is accurate and specific. Reasoning correctly eliminates alternatives. Minor weakness: could describe acoustic properties in more detail.}
\end{tcolorbox}

\bigskip

\begin{tcolorbox}[perturbbox, colback=gray!10, colframe=gray!50, title=Score 5 -- Excellent]
\textbf{[PERCEPTION]} Entities: acoustic guitar, finger plucking, string resonance.\\
Description: The audio features repeated finger-plucked notes on a string instrument, with a warm mid-range tone, subtle string buzz, and natural decay characteristic of an acoustic guitar.

\medskip
\textbf{[ANALYSIS]} The finger-plucking technique and warm mid-range resonance are hallmarks of an acoustic guitar. The natural decay and subtle string buzz are physical properties of nylon or steel strings, absent in piano (hammered), violin (bowed), or trumpet (air column vibration).

\medskip
\textbf{[REASONING]} The combination of plucking technique, warm resonance, string buzz, and natural decay conclusively identifies the instrument as a guitar. All other options are eliminated based on their distinct production mechanisms.

\medskip
\textbf{[ANSWER]} C. Guitar
\tcblower
\textit{Perception is detailed and grounded in specific acoustic properties. Analysis correctly identifies distinguishing features and eliminates all alternatives with clear justification. Reasoning is coherent and leads directly to the correct answer.}
\end{tcolorbox}

\section{M1 rational}
\label{app:M1}

The goal of M1 is not to capture all aspects of audio grounding but serves as a fundamental perception-level check: it evaluates whether the model correctly identifies relevant audio entities before performing further reasoning. Without accurate audio entity recognition, subsequent reasoning may be based on hallucinated or unsupported assumptions, even if the final answer happens to be correct.

To further examine the concern about generic audio-related terms, we conducted an analysis on MMAR. The frequencies of common generic terms (including ``speech'', ``voice'', ``music'', ``song'', ``instrument'', ``audio'', and ``sound'') in CoT generated across all models. The results are shown in Table \ref{tab:generic_tokens_m1}. 

\begin{table}[h]
\centering
\caption{Number of generic tokens and corresponding M1 scores across models.}
\small
\begin{tabular}{lcc}
\toprule
\textbf{Model} & \textbf{Generic Tokens} $\downarrow$ & \textbf{M1 Score} $\uparrow$ \\
\midrule
Qwen2-Audio      & 133   & 0.1963 \\
Qwen2.5-Omni     & 3,241 & 0.4344 \\
Gemma-3n-E2B     & 117   & 0.2861 \\
R1-AQA           & 117   & 0.2107 \\
Audio-Reasoner   & 4,272 & 0.2401 \\
AudSemThinker    & 6,581 & 0.3309 \\
Phi-4            & 5,553 & 0.3432 \\
GPT-4o-audio     & 9     & 0.3725 \\
Gemini 2.5 Flash & 650   & 0.3808 \\
\bottomrule
\end{tabular}

\label{tab:generic_tokens_m1}
\end{table}

Models with the highest frequency of generic audio-related words do not consistently achieve higher M1 scores as we require more accurate identification of the audio entities, also the generic terms does not necessarily lead to poor M1 scores, as these terms may still correspond to valid audio entities. This suggests that M1 is not primarily driven by superficial audio-related vocabulary frequency.

More specifically, we conducted an adversarial analysis using GPT-4o-audio outputs on MMAR, with results shown in Table \ref{tab:m1_validation}. 

\begin{table}[h]
\centering
\small
\caption{Validation of M1 under different entity manipulation strategies.}
\resizebox{\columnwidth}{!}{
\begin{tabular}{llc}
\toprule
\textbf{Mode} & \textbf{Description} & \textbf{M1 Mean} $\uparrow$ \\
\midrule
a. Original Perception
& Original CoT [Perception]
& \textbf{0.3949} \\

b. All ``music''
& Replace all entities with ``music''
& 0.1746 \\

c. All ``speech''
& Replace all entities with ``speech''
& 0.0940 \\

d. Replace half words
& Randomly replace half entities with ``music''/``speech'', keep the rest
& 0.3503 \\
\bottomrule
\end{tabular}}
\label{tab:m1_validation}
\end{table}

The results show that superficial manipulation cannot easily achieve a high M1 score, supporting the validity of M1 as an audio entity grounding measure. To get higher score we still need more accurate audio entity identification.

\end{document}